\documentclass[a4paper,11pt]{article}

\usepackage{jheppub} 

\usepackage[T1]{fontenc} 

\usepackage[dvipsnames]{xcolor}
\usepackage{slashed}
\usepackage[normalem]{ulem}

\usepackage{multirow}
\usepackage{comment}
\usepackage{overpic}
\graphicspath{{./figures/}}

\input{alias.tex}
\def\DzToKPiEta {D^{0}\to K^{-}\pi^{+}\eta}
\def\DzToPiPiEta {D^{0}\to\pi^{+}\pi^{-}\eta}
\def\DzToKKEta {D^{0}\to K^{+}K^{-}\eta}
\def\DzToPhiEta {D^{0}\to\phi\eta}

\newsavebox{\tablebox}

\title{\boldmath Measurement of branching fractions and search for $\CP$ violation in $\DzToPiPiEta$, $\DzToKKEta$, and $\DzToPhiEta$ at Belle}

\newcounter{AffiliationCounter}
\stepcounter{AffiliationCounter}\edef\instBilbao{\protect\theAffiliationCounter}
\stepcounter{AffiliationCounter}\edef\instBonn{\protect\theAffiliationCounter}
\stepcounter{AffiliationCounter}\edef\instBNL{\protect\theAffiliationCounter}
\stepcounter{AffiliationCounter}\edef\instBINP{\protect\theAffiliationCounter}
\stepcounter{AffiliationCounter}\edef\instCharles{\protect\theAffiliationCounter}
\stepcounter{AffiliationCounter}\edef\instChonnam{\protect\theAffiliationCounter}
\stepcounter{AffiliationCounter}\edef\instCincinnati{\protect\theAffiliationCounter}
\stepcounter{AffiliationCounter}\edef\instDESY{\protect\theAffiliationCounter}
\stepcounter{AffiliationCounter}\edef\instFlorida{\protect\theAffiliationCounter}
\stepcounter{AffiliationCounter}\edef\instFuJen{\protect\theAffiliationCounter}
\stepcounter{AffiliationCounter}\edef\instFudan{\protect\theAffiliationCounter}
\stepcounter{AffiliationCounter}\edef\instGifu{\protect\theAffiliationCounter}
\stepcounter{AffiliationCounter}\edef\instSokendai{\protect\theAffiliationCounter}
\stepcounter{AffiliationCounter}\edef\instGyeongsang{\protect\theAffiliationCounter}
\stepcounter{AffiliationCounter}\edef\instHanyang{\protect\theAffiliationCounter}
\stepcounter{AffiliationCounter}\edef\instHawaii{\protect\theAffiliationCounter}
\stepcounter{AffiliationCounter}\edef\instKEK{\protect\theAffiliationCounter}
\stepcounter{AffiliationCounter}\edef\instJPARC{\protect\theAffiliationCounter}
\stepcounter{AffiliationCounter}\edef\instHSE{\protect\theAffiliationCounter}
\stepcounter{AffiliationCounter}\edef\instJuelich{\protect\theAffiliationCounter}
\stepcounter{AffiliationCounter}\edef\instIKER{\protect\theAffiliationCounter}
\stepcounter{AffiliationCounter}\edef\instIITB{\protect\theAffiliationCounter}
\stepcounter{AffiliationCounter}\edef\instIITH{\protect\theAffiliationCounter}
\stepcounter{AffiliationCounter}\edef\instIITM{\protect\theAffiliationCounter}
\stepcounter{AffiliationCounter}\edef\instIndiana{\protect\theAffiliationCounter}
\stepcounter{AffiliationCounter}\edef\instIHEP{\protect\theAffiliationCounter}
\stepcounter{AffiliationCounter}\edef\instProtvino{\protect\theAffiliationCounter}
\stepcounter{AffiliationCounter}\edef\instVienna{\protect\theAffiliationCounter}
\stepcounter{AffiliationCounter}\edef\instNapoli{\protect\theAffiliationCounter}
\stepcounter{AffiliationCounter}\edef\instRomaTre{\protect\theAffiliationCounter}
\stepcounter{AffiliationCounter}\edef\instJAEA{\protect\theAffiliationCounter}
\stepcounter{AffiliationCounter}\edef\instJSI{\protect\theAffiliationCounter}
\stepcounter{AffiliationCounter}\edef\instKarlsruhe{\protect\theAffiliationCounter}
\stepcounter{AffiliationCounter}\edef\instKAU{\protect\theAffiliationCounter}
\stepcounter{AffiliationCounter}\edef\instKitasato{\protect\theAffiliationCounter}
\stepcounter{AffiliationCounter}\edef\instKISTI{\protect\theAffiliationCounter}
\stepcounter{AffiliationCounter}\edef\instKorea{\protect\theAffiliationCounter}
\stepcounter{AffiliationCounter}\edef\instKyotoSangyo{\protect\theAffiliationCounter}
\stepcounter{AffiliationCounter}\edef\instKyungpook{\protect\theAffiliationCounter}
\stepcounter{AffiliationCounter}\edef\instLAL{\protect\theAffiliationCounter}
\stepcounter{AffiliationCounter}\edef\instLebedev{\protect\theAffiliationCounter}
\stepcounter{AffiliationCounter}\edef\instLNNU{\protect\theAffiliationCounter}
\stepcounter{AffiliationCounter}\edef\instLjubljana{\protect\theAffiliationCounter}
\stepcounter{AffiliationCounter}\edef\instLMU{\protect\theAffiliationCounter}
\stepcounter{AffiliationCounter}\edef\instLuther{\protect\theAffiliationCounter}
\stepcounter{AffiliationCounter}\edef\instMNIT{\protect\theAffiliationCounter}
\stepcounter{AffiliationCounter}\edef\instMaribor{\protect\theAffiliationCounter}
\stepcounter{AffiliationCounter}\edef\instMPI{\protect\theAffiliationCounter}
\stepcounter{AffiliationCounter}\edef\instMississippi{\protect\theAffiliationCounter}
\stepcounter{AffiliationCounter}\edef\instMEPhI{\protect\theAffiliationCounter}
\stepcounter{AffiliationCounter}\edef\instNagoya{\protect\theAffiliationCounter}
\stepcounter{AffiliationCounter}\edef\instUNapoli{\protect\theAffiliationCounter}
\stepcounter{AffiliationCounter}\edef\instNara{\protect\theAffiliationCounter}
\stepcounter{AffiliationCounter}\edef\instNCU{\protect\theAffiliationCounter}
\stepcounter{AffiliationCounter}\edef\instNUU{\protect\theAffiliationCounter}
\stepcounter{AffiliationCounter}\edef\instTaiwan{\protect\theAffiliationCounter}
\stepcounter{AffiliationCounter}\edef\instKrakow{\protect\theAffiliationCounter}
\stepcounter{AffiliationCounter}\edef\instNihonDental{\protect\theAffiliationCounter}
\stepcounter{AffiliationCounter}\edef\instNiigata{\protect\theAffiliationCounter}
\stepcounter{AffiliationCounter}\edef\instNovosibirsk{\protect\theAffiliationCounter}
\stepcounter{AffiliationCounter}\edef\instOsakaCity{\protect\theAffiliationCounter}
\stepcounter{AffiliationCounter}\edef\instPNNL{\protect\theAffiliationCounter}
\stepcounter{AffiliationCounter}\edef\instPanjab{\protect\theAffiliationCounter}
\stepcounter{AffiliationCounter}\edef\instPittsburgh{\protect\theAffiliationCounter}
\stepcounter{AffiliationCounter}\edef\instNPC{\protect\theAffiliationCounter}
\stepcounter{AffiliationCounter}\edef\instRIKENMSL{\protect\theAffiliationCounter}
\stepcounter{AffiliationCounter}\edef\instURomaTre{\protect\theAffiliationCounter}
\stepcounter{AffiliationCounter}\edef\instUSTC{\protect\theAffiliationCounter}
\stepcounter{AffiliationCounter}\edef\instSeoul{\protect\theAffiliationCounter}
\stepcounter{AffiliationCounter}\edef\instShoyaku{\protect\theAffiliationCounter}
\stepcounter{AffiliationCounter}\edef\instSoongsil{\protect\theAffiliationCounter}
\stepcounter{AffiliationCounter}\edef\instSungkyunkwan{\protect\theAffiliationCounter}
\stepcounter{AffiliationCounter}\edef\instSydney{\protect\theAffiliationCounter}
\stepcounter{AffiliationCounter}\edef\instTabuk{\protect\theAffiliationCounter}
\stepcounter{AffiliationCounter}\edef\instTata{\protect\theAffiliationCounter}
\stepcounter{AffiliationCounter}\edef\instTUM{\protect\theAffiliationCounter}
\stepcounter{AffiliationCounter}\edef\instTelAviv{\protect\theAffiliationCounter}
\stepcounter{AffiliationCounter}\edef\instTohoku{\protect\theAffiliationCounter}
\stepcounter{AffiliationCounter}\edef\instERI{\protect\theAffiliationCounter}
\stepcounter{AffiliationCounter}\edef\instTokyo{\protect\theAffiliationCounter}
\stepcounter{AffiliationCounter}\edef\instTIT{\protect\theAffiliationCounter}
\stepcounter{AffiliationCounter}\edef\instTMU{\protect\theAffiliationCounter}
\stepcounter{AffiliationCounter}\edef\instVPI{\protect\theAffiliationCounter}
\stepcounter{AffiliationCounter}\edef\instWayneState{\protect\theAffiliationCounter}
\stepcounter{AffiliationCounter}\edef\instYamagata{\protect\theAffiliationCounter}
\stepcounter{AffiliationCounter}\edef\instYonsei{\protect\theAffiliationCounter}

\collaboration{The BELLE collaboration}
  \author[\instCincinnati]{L.~K.~Li,} 
  \author[\instCincinnati]{A.~J.~Schwartz,} 
  \author[\instCincinnati]{K.~Kinoshita,} 

  \author[\instKEK,\instSokendai]{I.~Adachi,} 
  \author[\instTokyo]{H.~Aihara,} 
  \author[\instTabuk,\instKAU]{S.~Al~Said,} 
  \author[\instBNL]{D.~M.~Asner,} 
  \author[\instCincinnati]{H.~Atmacan,} 
  \author[\instBINP,\instNovosibirsk]{V.~Aulchenko,} 
  \author[\instHSE]{T.~Aushev,} 
% ...

\emailAdd{lilk@ucmail.uc.edu}

\arxivnumber{2106.04286}

\abstract{
We measure the branching fractions and $\CP$ asymmetries for the singly 
Cabibbo-suppressed decays $\DzToPiPiEta$, $\DzToKKEta$, and $\DzToPhiEta$,t
using 980~fb$^{-1}$ of data from the Belle experiment at the 
KEKB $e^+e^-$ collider.
We obtain
\begin{eqnarray}
\mathcal{B}(\DzToPiPiEta)  & = &   
[1.22\pm 0.02\,({\rm stat})\pm 0.02\,({\rm syst})\pm 0.03\,(\mathcal{B}_{\rm ref})]\times 10^{-3}\,, \nonumber \\ 
\mathcal{B}(\DzToKKEta)  & = &  
[1.80\,^{+0.07}_{-0.06}\,({\rm stat})\pm 0.04\,({\rm syst})\pm 0.05\,(\mathcal{B}_{\rm ref})]\times 10^{-4}\,, \nonumber \\ 
\mathcal{B}(\DzToPhiEta) & = & 
[1.84\pm 0.09\,({\rm stat})\pm 0.06\,({\rm syst})\pm 0.05\,(\mathcal{B}_{\rm ref})]\times 10^{-4}\,, \nonumber
 \end{eqnarray}
where the third uncertainty ($\mathcal{B}_{\rm ref}$) is from the uncertainty in the branching fraction of the reference mode $\DzToKPiEta$. The color-suppressed decay $\DzToPhiEta$ is observed for the first time, with 
very high significance.
The results for the $\CP$ asymmetries are
\begin{eqnarray}
\Acp(\DzToPiPiEta) & = & [0.9\pm 1.2\,({\rm stat})\pm 0.5\,({\rm syst})]\%\,, \nonumber \\
\Acp(\DzToKKEta)   & = & [-1.4\pm 3.3\,({\rm stat})\pm 1.1\,({\rm syst})]\%\,, \nonumber \\
\Acp(\Dz\to\phi\eta)&= & [-1.9\pm 4.4\,({\rm stat})\pm 0.6\,({\rm syst})]\%\,. \nonumber
\end{eqnarray}
The results for $\DzToPiPiEta$ are a significant improvement over
previous results. 
The branching fraction and $\Acp$ results for $\DzToKKEta$, and 
the $\Acp$ result for $\DzToPhiEta$,
are the first such measurements. 
No evidence for $\CP$ violation is found in any of these decays.
}

\begin{document}


\preprint{\vbox{ 
\hbox{Belle Preprint 2021-11}
\hbox{KEK Preprint 2021-8}
}}
\maketitle
\flushbottom 


\section{\boldmath Introduction}
Singly Cabibbo-suppressed (SCS) decays of charmed mesons provide a
promising opportunity to study $\CP$ violation in the charm sector.
Within the Standard Model, $\CP$ violation in charm 
decays is expected to be of the order of 
$10^{-3}$ or smaller~\cite{bib:PRD85o034036,bib:PRD75d036008}, 
and thus challenging to observe.
SCS decays are of special interest, as interference 
that includes a new physics amplitude could lead 
to large $\CP$ violation. 
The $\CP$ asymmetry between $D^0\to f$ and 
$\Dzb\to\bar{f}$ decays ($\Acp$) is defined as
\begin{eqnarray}
\Acp & = & 
\frac{\mathcal{B}(\Dz\to f)-\mathcal{B}(\Dzb\to \bar{f})}
      {\mathcal{B}(\Dz\to f) + \mathcal{B}(\Dzb\to \bar{f})}\,.
      \label{eqn:acp}
\end{eqnarray}
The only observation of $\CP$ violation in the charm sector to date is from the LHCb experiment, where a difference in $\Acp$ between the SCS $D^0\to\Kp\Km$ and $D^0\to\pip\pim$ decays~\cite{bib:CPVobservation} was observed: $\Delta\Acp=(-15.4\pm 2.9)\times10^{-4}$.
In this paper, we investigate two analogous
SCS decays, $\DzToPiPiEta$ and $\DzToKKEta$.
A search for a $\CP$ asymmetry in the first decay was performed by the BESIII experiment; the resulting precision was~6\%~\cite{bib:PRD101d052009}. 
There have been no results for $\DzToKKEta$ decays to date.
Theoretically, it is difficult to predict $\CP$ asymmetries for three-body decays, 
while some predictions exist for intermediate two-body processes:
$\Acp(\Dz\to\rho^0\eta)$ is predicted to be $-0.53\times10^{-3}$ from
tree amplitudes alone, and $-0.23\times 10^{-3}$ after considering QCD-penguin and weak penguin-annihilation~\cite{bib:PRD85o034036}. The asymmetry $\Acp(\Dz\to\phi\eta)$ is predicted to be zero in several theoretical models~\cite{bib:PRD85o034036}. 
A precise measurement of branching fractions 
($\mathcal{B}$) for these three-body decays is an 
important step towards searching for $\CP$ violation in these channels. 

In this paper we utilize the full Belle data sample of 980~$\invfb$ 
to measure $\mathcal{B}$ and $\Acp$
for three SCS decays: $\DzToPiPiEta$, $\DzToKKEta$, and $\DzToPhiEta$.
All $\mathcal{B}$ measurements are performed relative to the 
Cabibbo-favored (CF) decay $\DzToKPiEta$, which has been 
well-measured (with a fractional uncertainty 
$\delta\mathcal{B}/\mathcal{B}\sim3\%$~\cite{bib:PDG2020}) by both Belle~\cite{bib:PRD102d012002} and BESIII~\cite{bib:PRL124d241803}.
The current world average for $\mathcal{B}(\DzToPiPiEta)$
has a fractional uncertainty $\delta\mathcal{B}/\mathcal{B}\sim6\%$~\cite{bib:PDG2020}.
The branching fraction for $\DzToPhiEta$ was previously 
measured by Belle with 78~$\invfb$ of data~\cite{bib:PRL92d101803}; the measurement 
reported here uses an order of magnitude more data and supersedes 
that result. 
BESIII found evidence for $\DzToPhiEta$ ($4.2\sigma$)~\cite{bib:PLB798d135017} and observed a non-$\phi$ $\DzToKKEta$ component ($5.2\sigma$)~\cite{bib:PRL124d241803}.

To identify the flavor of the neutral $D$
meson when produced, we reconstruct $D^{*+}\to\Dz\pi^+_s$ and
$D^{*-}\to\Dzb\pi^-_s$ decays; 
the charge of the daughter $\pi^\pm_s$ (which has low momentum and is
referred to as the ``slow'' pion) identifies whether the $D$ 
meson is $\Dz$ or $\Dzb$.
The raw asymmetry measured ($A_{\rm raw}$)
receives contributions from several sources:
\begin{eqnarray}
A_{\rm raw} & = & A_{\CP}^{\Dz\to f} + A_{\rm FB}^{D^{*+}} + A_{\varepsilon}^{\pi_{s}}\,, 
 \end{eqnarray}
where $A_{\CP}^{\Dz\to f}$ is the $\CP$ asymmetry for $\Dz\to f$;
$A_{\rm FB}^{D^{*+}}$ is the forward-backward asymmetry due to
$\gamma$-$Z^0$ interference and higher-order QED effects~\cite{bib:BROWN1973403}
in $e^+e^-\ra c\bar{c}$ collisions;
and $A_{\varepsilon}^{\pi_s}$ is the asymmetry resulting from a difference
in reconstruction efficiencies between $\pi^+_s$ and $\pi^-_s$.
This asymmetry depends on the transverse momentum $p_{T}(\pi_s)$ and 
polar angle $\theta(\pi_s)$ of the $\pi_s$ in the laboratory frame.
We correct for this by weighting signal events by a factor
$[1 - A_{\varepsilon}^{\pi_s}(p_T, \cos\theta)]$
for $\Dz$ decays, and by a factor
$[1 + A_{\varepsilon}^{\pi_s}(p_T, \cos\theta)]$
for $\Dzb$ decays.
After this weighting,
we are left with the $\pi_s$-corrected asymmetry
\begin{eqnarray}
A_{\rm corr}(\cos\theta^{*})  & = & \Acp + A_{\rm FB}(\cos\theta^{*})\,.
\end{eqnarray}
Since $A_{\rm FB}$ is an odd function of 
the cosine of the $D^{*+}$ polar angle $\theta^{*}$ in the $e^+e^-$ center-of-mass (CM) frame, and $A_{\CP}$ is independent of $\cos\theta^{*}$, we extract $\Acp$ and $A_{\rm FB}(\cos\theta^{*})$ via
\begin{eqnarray}
\Acp & = & 
\dfrac{ A_{\rm corr}(\cos\theta^{*}) +  A_{\rm corr}(-\cos\theta^{*})}{2}\,, \label{eqn:Acp1} \\
A_{\rm FB} (\cos\theta^{*}) & = & 
\dfrac{ A_{\rm corr}(\cos\theta^{*})- A_{\rm corr}(-\cos\theta^{*})}{2}\,. \label{eqn:Acp2}
\end{eqnarray}
Fitting the values of $\Acp$ for different $\cos\theta^*$ bins to a constant gives our final measurement of $\Acp$ for $\Dz\to f$.

\section{Belle detector and data sets}\label{sec:data}
This measurement is based on the full data set of the Belle experiment, 
which corresponds to a total integrated luminosity of 980~$\invfb$~\cite{bib:BelleDetector2} collected at or 
near the $\Upsilon(nS)$ ($n=1$, 2, 3, 4, 5) resonances.
The Belle experiment ran at the 
KEKB energy-asymmetric collider~\cite{bib:KEKB,bib:KEKB2}.
The Belle detector is a large-solid-angle magnetic spectrometer 
consisting of a silicon vertex detector (SVD), a $50$-layer central drift chamber (CDC), an array of aerogel threshold Cherenkov counters (ACC), a barrel-like arrangement of time-of-flight scintillation counters (TOF), and an electromagnetic calorimeter comprising CsI(Tl) crystals located inside a superconducting solenoid coil providing a $1.5$~T magnetic field. 
An iron flux-return located outside the coil is instrumented to detect 
$K_L^0$ mesons and to identify muons. A detailed description of the 
detector is given in Refs.~\cite{bib:BelleDetector,bib:BelleDetector2}. 

We use Monte Carlo (MC) simulated events to optimize selection criteria,
study backgrounds, and evaluate the signal reconstruction efficiency.
Signal MC events are generated by EVTGEN~\cite{bib:evtgen} and propagated through a detector simulation based on GEANT3~\cite{bib:geant3}. Final-state radiation from charged particles is simulated using the PHOTOS package~\cite{bib:PHOTOS}. Three-body decays are generated 
according to phase space.
An MC sample of ``generic'' events, 
corresponding to an integrated luminosity four times that of the data, 
is used to develop selection criteria. It includes 
$B\overline{B}$ events and continuum processes 
$\epem\to q\bar{q}$, where $q=u,d,s,c$. At the
$\Upsilon(5S)$ resonance, the MC includes 
$B^{(*)0}_{s}\,\overline{B}{}^{(*)0}_s$ events. 
Selection criteria are optimized by maximizing a 
figure-of-merit $N_{\rm sig}/\sqrt{N_{\rm sig}+N_{\rm bkg}}$, 
where $N_{\rm sig}$ and $N_{\rm bkg}$ are the numbers of signal 
and background events, respectively, expected in a 
two-dimensional signal region in variables $M$ and $Q$. 
The variable $M$ is the invariant mass of the 
$h^+h^-\eta$ ($h=\pi,K$) combination, and $Q=[M(h^+h^-\eta\,\pis)-M(h^+h^-\eta)-m_{\pis}]\cdot c^{2}$ is the
kinetic energy released in the $D^{*+}$ decay.

\section{Event selection and optimization}
\label{sec:selection}
We reconstruct the signal decays $\DzToPiPiEta$ and $\DzToKKEta$, and 
the reference decay $\DzToKPiEta$, in which the $\Dz$ originates from
$D^{*+}\to\Dz\pi^+$, as follows.\footnote{Throughout this paper,
charge-conjugate modes are implicitly included unless stated otherwise.}
Charged tracks are identified as $K^{\pm}$ or $\pi^{\pm}$ candidates 
using a likelihood ratio
$\mathcal{R}_K\equiv\mathcal{L}_K/(\mathcal{L}_K+\mathcal{L}_\pi)$,
where $\mathcal{L}_{K}$ ($\mathcal{L}_\pi$) is the likelihood
that a track is a $K^{\pm}$ ($\pi^{\pm}$) based on the photon yield in the ACC, $dE/dx$ information in the CDC, and time-of-flight information from the TOF~\cite{bib:PID}. 
Tracks having $\mathcal{R}_{K}>0.60$ are identified as $K^{\pm}$ 
candidates; otherwise, they are considered as $\pi^{\pm}$ candidates. 
The corresponding efficiencies are approximately $90\%$ 
for kaons and $95\%$ for pions.
Tracks that are highly electron-like ($\mathcal{R}_e>0.95$) or muon-like ($\mathcal{R}_\mu>0.95$) are rejected, where the electron and muon likelihood ratios $\mathcal{R}_e$ and $\mathcal{R}_\mu$ are
determined mainly using information from the ECL and KLM detectors, respectively~\cite{bib:NIMA485d490,bib:NIMA491d69}.
Charged tracks are required to have at least two SVD hits 
in the $+z$ direction (defined as the direction opposite that 
of the positron beam), and at least two SVD hits in the 
$x$-$y$ (transverse) plane. 
The nearest approach of the $\pi_s^+$ track to the 
$e^+e^-$ interaction point (IP) is required to be less 
than 1.0~cm in the $x$-$y$ plane, and 
less than 3.0~cm along the $z$ axis.

Photon candidates are identified as energy clusters in the ECL that 
are not associated with any charged track. The photon energy
($E_{\gamma}$) is required to be greater than 50~MeV in 
the barrel region (covering the polar angle $32^{\circ}<\theta<129^{\circ}$), and greater than 100~MeV in the endcap region ($12^{\circ}<\theta<31^{\circ}$ or $132^{\circ}<\theta<157^{\circ}$). 
The ratio of
the energy deposited in the $3\times 3$ array of crystals centered
on the crystal with the highest energy, to the energy deposited in 
the corresponding $5\times 5$ array of crystals, is required to be 
greater than~0.80.

Candidate $\eta\to\gamma\gamma$ decays are reconstructed 
from photon pairs having an invariant mass satisfying
$500~{\rm MeV}/c^2 < M(\gamma\gamma) < 580~{\rm MeV}/c^2$.
This range corresponds to about
$3\sigma$ in $M(\gamma\gamma)$ resolution.
The absolute value of the cosine of the
$\eta\to\gamma_{1}\gamma_{2}$ decay angle, defined as
$\cos\theta_{\eta}\equiv
E(\eta)/p(\eta)\cdot 
(E_{\gamma_1}-E_{\gamma_2})/(E_{\gamma_1}+E_{\gamma_2})$,
is required to be less than~0.85. This 
retains around 89\% of the signal 
while reducing backgrounds by a factor of two.
To further suppress backgrounds,
we remove $\eta$ candidates in which both photon daughters
can be combined with other photons in the event to form $\piz\to\gamma\gamma$ candidate decays satisfying 
$|M_{\gamma\gamma}-m_{\piz}|<10$~MeV/$c^2$, where $m^{}_{\pi^0}$ is the nominal $\pi^0$ mass~\cite{bib:PDG2020}. This veto requirement has an
efficiency of 95\% while reducing backgrounds by 
a factor of three ($\DzToKKEta$) and four ($\DzToPiPiEta$).

Candidate $\DzToPiPiEta$, $\DzToKKEta$, and $\DzToKPiEta$
decays are reconstructed by combining $\pi^\pm$ and $K^\pm$ 
tracks with $\eta$ candidates.
A vertex fit is performed with the two charged tracks to obtain
the $D^0$ decay vertex position; the resulting fit quality is 
labeled~$\chi^2_{v}$. 
To improve the momentum resolution of the $\eta$, 
the $\gamma$ daughters are subjected to a fit 
in which the photons are required to originate from the 
$D^{0}$ vertex position, and the invariant mass is 
constrained to be that of the $\eta$ meson~\cite{bib:PDG2020}.
The fit quality of this mass constraint ($\chi_{m}^{2}$) is required to satisfy~$\chi_m^2<8$, 
and the resulting $\eta$ momentum is 
required to be greater than 0.70~GeV/$c$.
For $\DzToPiPiEta$ candidates, we veto events in which
$|M(\pip\pim)-m_{\KS}|<10$~MeV/$c^2$, where $m^{}_{\KS}$ 
is the nominal $\KS$ mass~\cite{bib:PDG2020}, to suppress 
background from CF $\Dz\to\KS\,\eta$ decays. This 
veto range corresponds to about $3\sigma$ in resolution.
The $D^0$ invariant mass $M$ is required to satisfy
$1.850~{\rm GeV}/c^2 < M < 1.878~{\rm GeV}/c^2$
for $\DzToKKEta$ candidates;
$1.840~{\rm GeV}/c^2 < M < 1.884~{\rm GeV}/c^2$
for $\DzToPiPiEta$ candidates; and 
$1.842~{\rm GeV}/c^2 < M < 1.882~{\rm GeV}/c^2$
for $\DzToKPiEta$ candidates.  
These ranges correspond to about $2\sigma$ in resolution. 

Candidate $D^{*+}\to\Dz\pi_s^+$ decays are reconstructed 
by combining $D^0$ candidates with $\pi_s^+$ tracks. 
We first fit for $D^{*+}$ decay vertex using the
$D^0$ momentum vector and decay vertex position, and the IP
as a constraint (i.e., the $D^{*+}$ nominally originates from the IP).
The resulting goodness-of-fit is labeled~$\chi^2_{\rm IP}$.
To improve the resolution in $Q$, another vertex fit 
is performed: in this case we constrain the $\pi_s^+$ 
daughter to originate from the
$D^{*+}$ decay vertex, and
the resulting fit quality is labeled~$\chi^2_{s}$. 
The sum of the above three fit qualities, 
$\sum\chi^2_{\rm vtx}=\chi^2_{v}+\chi^2_{\rm IP}+\chi^2_{s}$,
is required to be less than~50;
this requirement has a signal efficiency of about 97\%. 
Those $D^{*+}$ candidates satisfying $0<Q<15$~MeV 
are retained for further analysis.
To eliminate $D^{*+}$ candidates originating from $B$ decays, 
and to also suppress combinatorial background, the 
$D^{*+}$ momentum in the CM frame is required to 
be greater than 2.70~GeV/$c$.

After the above selection criteria are applied,
about 2.1\% of $\DzToPiPiEta$ events, 
1.3\% of $\DzToKPiEta$ events, and 
$<0.1$\% of $\DzToKKEta$ events 
have two or more $D^{*+}$ candidates.
For such multi-candidate events, we choose a single candidate:
that which has the smallest value of the sum 
$\sum\chi^2_{\rm vtx} + \chi_{m}^2(\eta)$.
This criterion,
according to MC simulation, identifies the 
correct candidate 54\% of the time.

\section{Measurement of the branching fractions}\label{sec:BF}

\subsection{\boldmath{Measurement of $\mathcal{B}(\DzToPiPiEta)$ 
and $\mathcal{B}(\DzToKKEta)$}}
\label{sec:BF1}
We extract the signal yield via an unbinned maximum-likelihood
fit to the $Q$ distribution. The probability density 
function (PDF) used for signal events is taken to be 
the sum of
a bifurcated Student's t-function ($S_{\rm bif}$), which is defined in appendix \ref{app:student}, 
and one or two asymmetric Gaussians ($G_{\rm asym}$), 
with all having a common mean. 
The PDF used for $\DzToKKEta$ signal events is simply an $S_{\rm bif}$ function. These PDFs are explicitly 
\begin{eqnarray}
\mathcal{P}_{\rm sig}^{K\pi\eta} & = &   f_1 [f_s S_{\rm bif}(\mu, \sigma_0, \delta_0, n_l, n_h)  + (1-f_s) G_{\rm asym}(\mu, r_1\sigma_0,\delta_1)]\nonumber \\
& &  + (1-f_1)G_{\rm asym}(\mu, r_2r_1\sigma_0,\delta_2)\,,
\label{eqn:qshape1} \\
\mathcal{P}_{\rm sig}^{\pi\pi\eta} & = &   f_s S_{\rm bif}(\mu, \sigma_0, \delta_0, n_l, n_h)  + (1-f_s) G_{\rm asym}(\mu, r_1\sigma_0,\delta_1)\,,
\label{eqn:qshape2} \\
\mathcal{P}_{\rm sig}^{KK\eta} & = &  S_{\rm bif}(\mu, \sigma_0, \delta_0, n_l, n_h)\,. \label{eqn:qshape3} 
\end{eqnarray}
In these expressions, 
$\delta_i$ is an asymmetry parameter characterizing the
difference between left-side and right-side widths: 
$\sigma^{}_{R,L}=\sigma(1\pm\delta)$.
Most of these parameters are fixed to values 
obtained from MC simulation. However, 
the parameters $\mu$, $\sigma_0$, and, for the higher-statistics
$\DzToKPiEta$ channel, $n_{l,h}$, are floated to account for possible
differences in resolution between data and MC.
For backgrounds, the PDF is taken to be a threshold function $f(Q)=Q^{\alpha} e^{-\beta Q}$;
for the  CF mode $\DzToKPiEta$, 
we include an additional symmetric Gaussian to describe 
a small background component originating from 
misreconstructed $\Dz\to\Km\pip\piz\piz$ decays.
The parameters of this Gaussian are fixed to values obtained 
from MC simulation, while all other parameters are floated.
No other peaking backgrounds, such as misreconstructed $\Dz$ decays or signal decays in which a pion from the $\Dz$ is swapped with that from the $D^{*+}$ decay, are found in the MC simulation.

\begin{figure}[tbp]
  \begin{center}
  \begin{overpic}[width=0.45\textwidth]{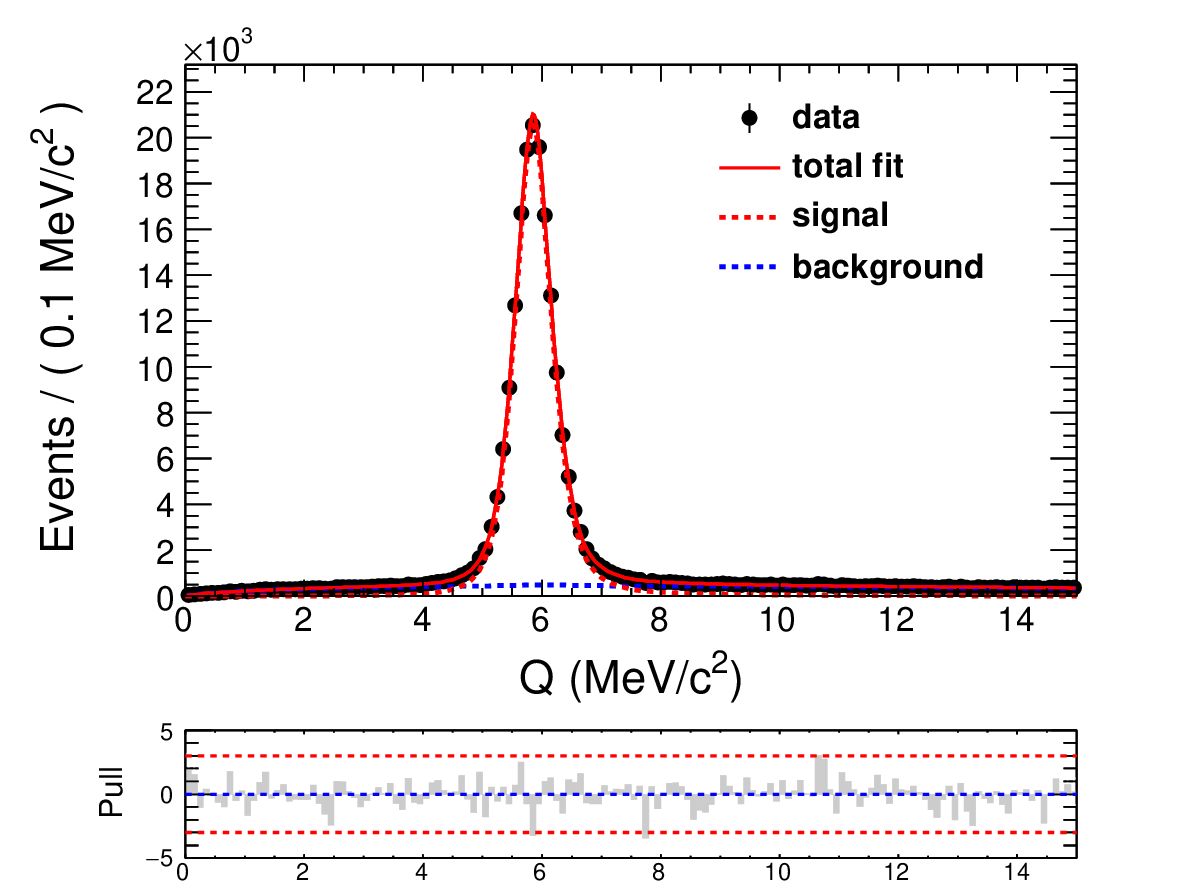}%
  \put(22,62){(a)}%
  \end{overpic}%
  \begin{overpic}[width=0.45\textwidth]{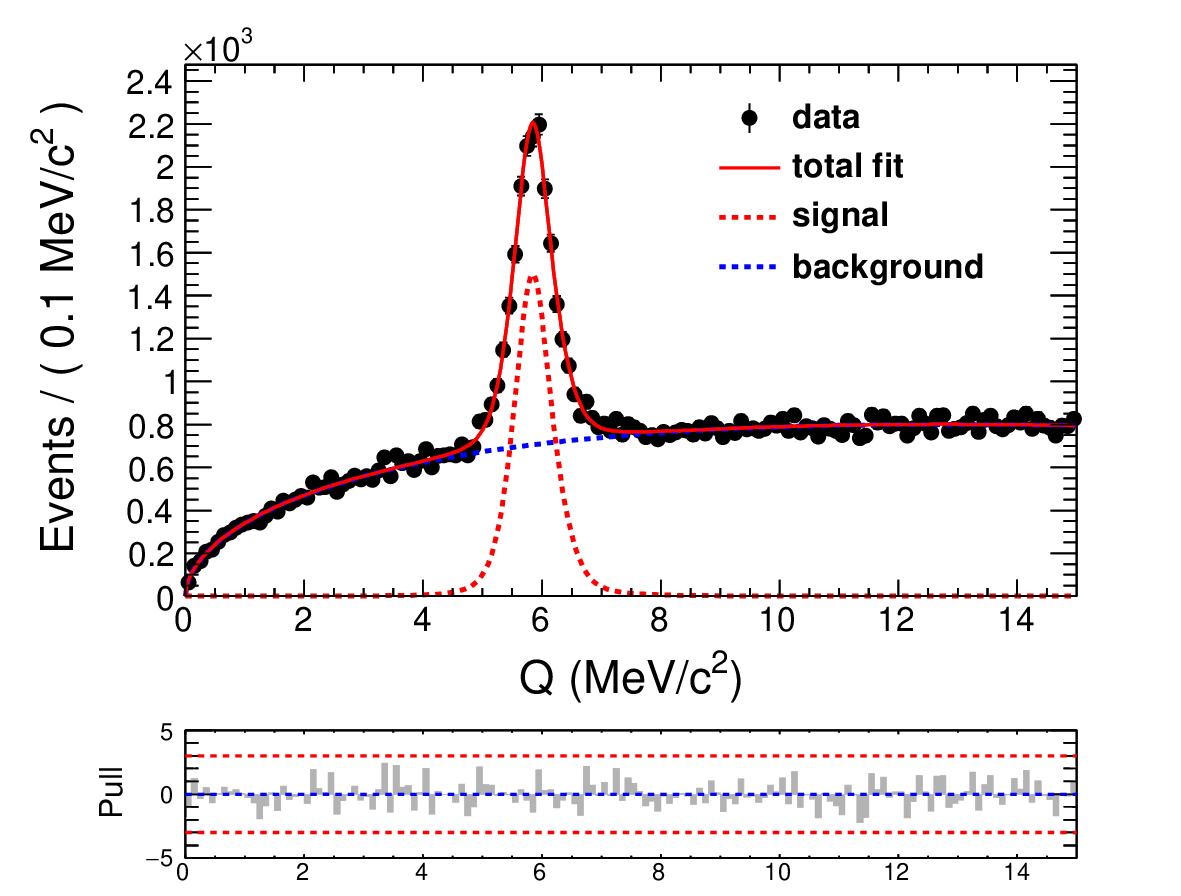}%
  \put(22,62){(b)}%
  \end{overpic}\\
  \begin{overpic}[width=0.45\textwidth]{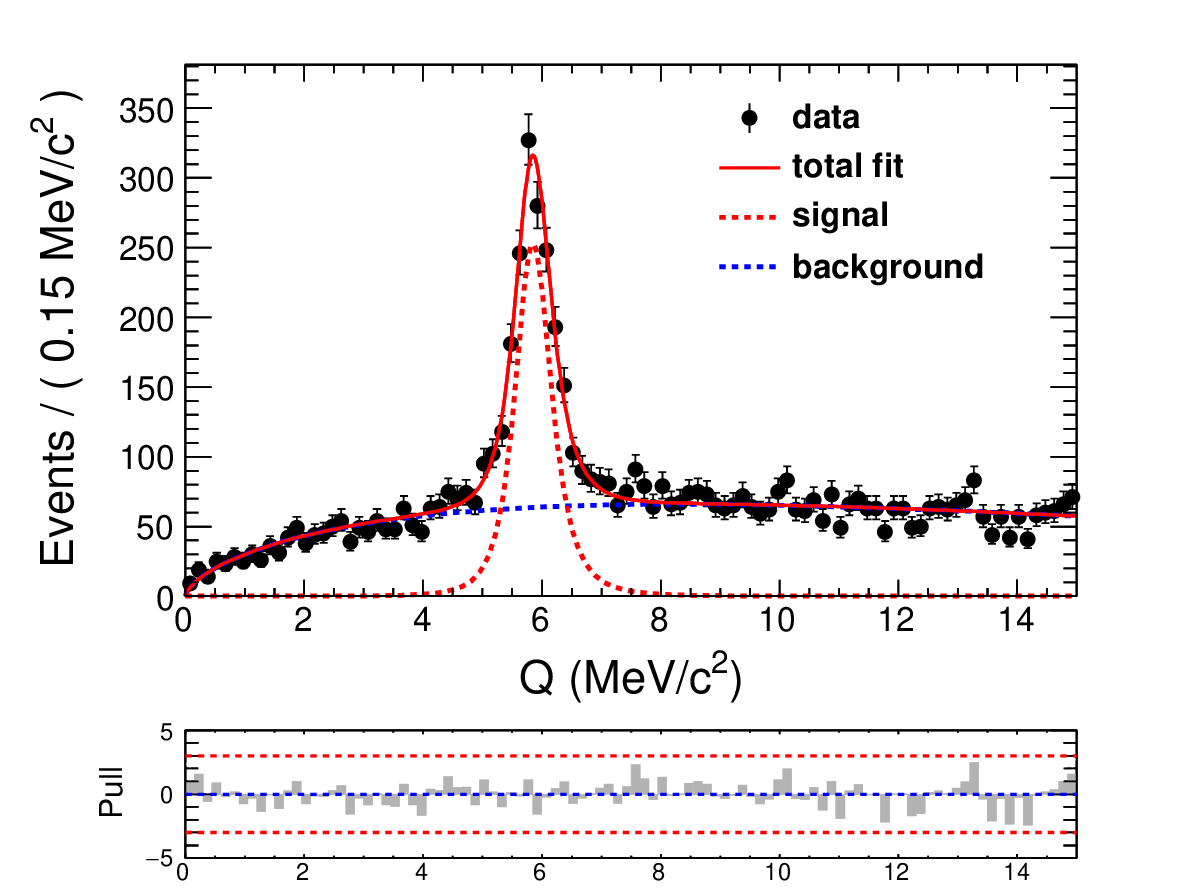}%
  \put(22,62){(c)}%
  \end{overpic}%
  \begin{overpic}[width=0.45\textwidth]{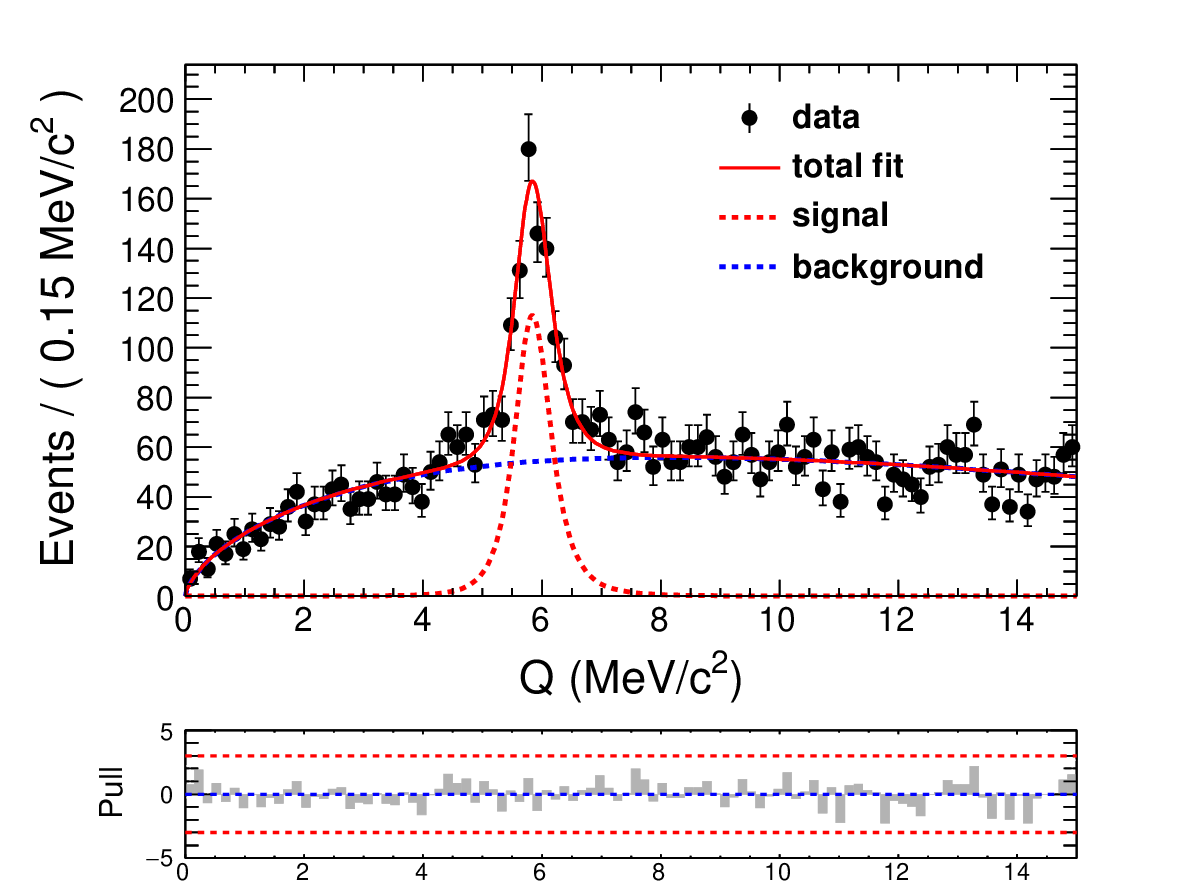}%
  \put(22,62){(d)}%
  \end{overpic}%
  \caption{Distributions of the released energy $Q$ in $D^{*+}\to\Dz\pis$ decay for (a) $\DzToKPiEta$, (b) $\DzToPiPiEta$, (c) $\DzToKKEta$, and (d) $\DzToKKEta$ with the $\phi$-peak excluded by requiring $|M_{KK}-m_\phi|>20$ MeV/$c^2$. Points with error bars show the data; the dashed red curve shows the signal; the dashed
  blue curve shows the background; and the solid red curves
  show the overall fit result. The pull plots underneath the fit results
  show the residuals divided by the errors in the histogram.}
  \label{fig:unBlindQ}
  \end{center}
\end{figure}

The results of the fit are shown in figure~\ref{fig:unBlindQ}, 
along with the pull 
$(N_{\rm data}-N_{\rm fit})/\sigma$, 
where $\sigma$ is the error on $N_{\rm data}$.
All fit residuals look satisfactory. The signal 
yields in the fitted region $0<Q<15$~MeV, and in the 
signal region $|Q-5.86|<0.80$~MeV, are listed in table~\ref{tabQ:yields_unblind2}. 

\begin{table}[!htbp]
\centering
\begin{tabular}{|c|c|c|c|c|} \hline 
Region & Component 	&  $\DzToKPiEta$ 	& $\DzToPiPiEta$ 		& $\DzToKKEta$		\\ \hline
\multirow{2}{*}{Fitted region} 
& signal			&  $180369 \pm 837$	& $12982 \pm 198$  		& $1482 \pm 60$ 	 	\\
& background		&  $57752 \pm 761$		& $101011 \pm 357$   	& $5681 \pm 88$ 	 	\\  \hline  
\multirow{2}{*}{Signal region} 
& signal			&  $162456 \pm 754$	& $12053 \pm 184$  	& $1343 \pm 54$ 	 	\\
& background		&  $7578 \pm 100$		& $11274 \pm 40$   	& $678 \pm 11$ 	 	\\    \hline
\end{tabular}
\caption{Yields of signal and background events in the 
fitted region $0<Q<15$~MeV, and in the signal region
$|Q-5.86|<0.80$~MeV.}
\label{tabQ:yields_unblind2}
\end{table} 

\begin{figure}[!hbtp]
  \begin{centering}
  \begin{overpic}[width=0.33\textwidth]{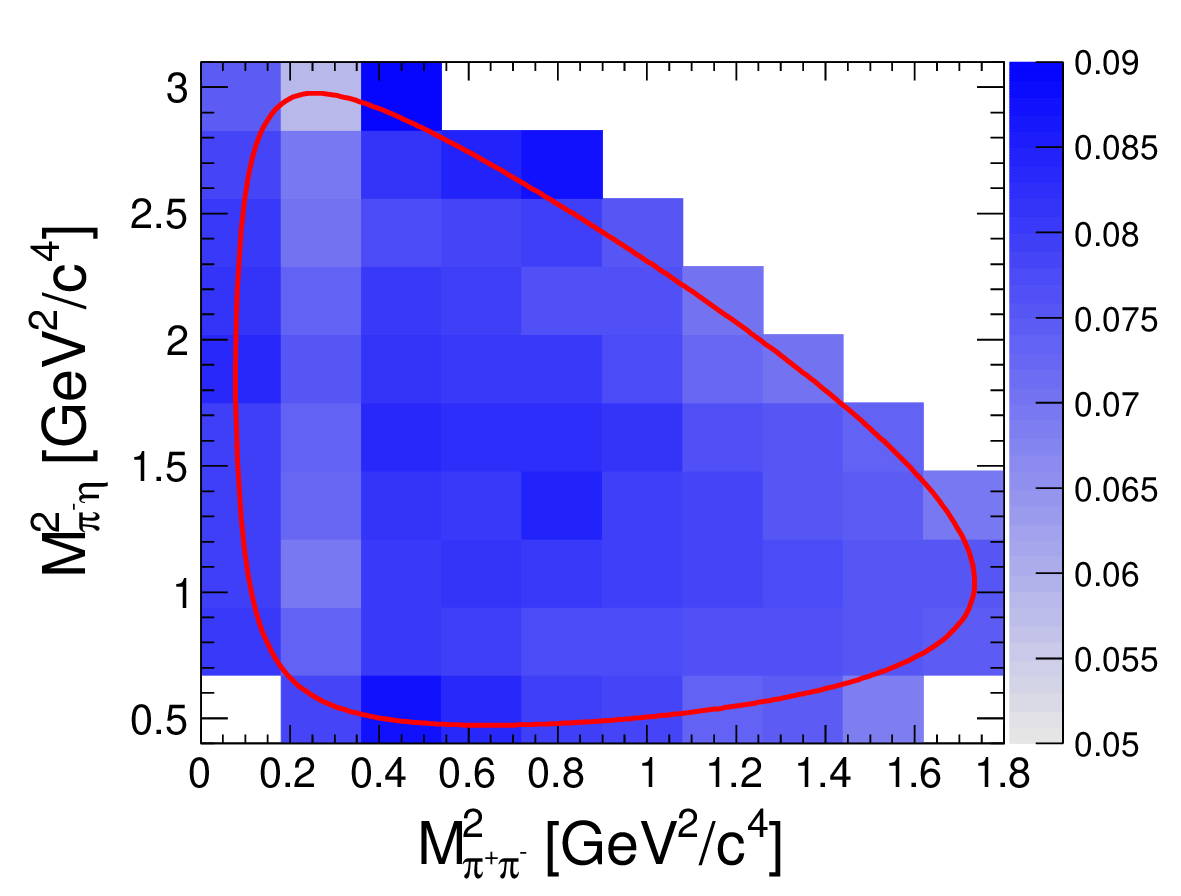}%
  \put(70,62){(a)}%
  \end{overpic}%
  \begin{overpic}[width=0.33\textwidth]{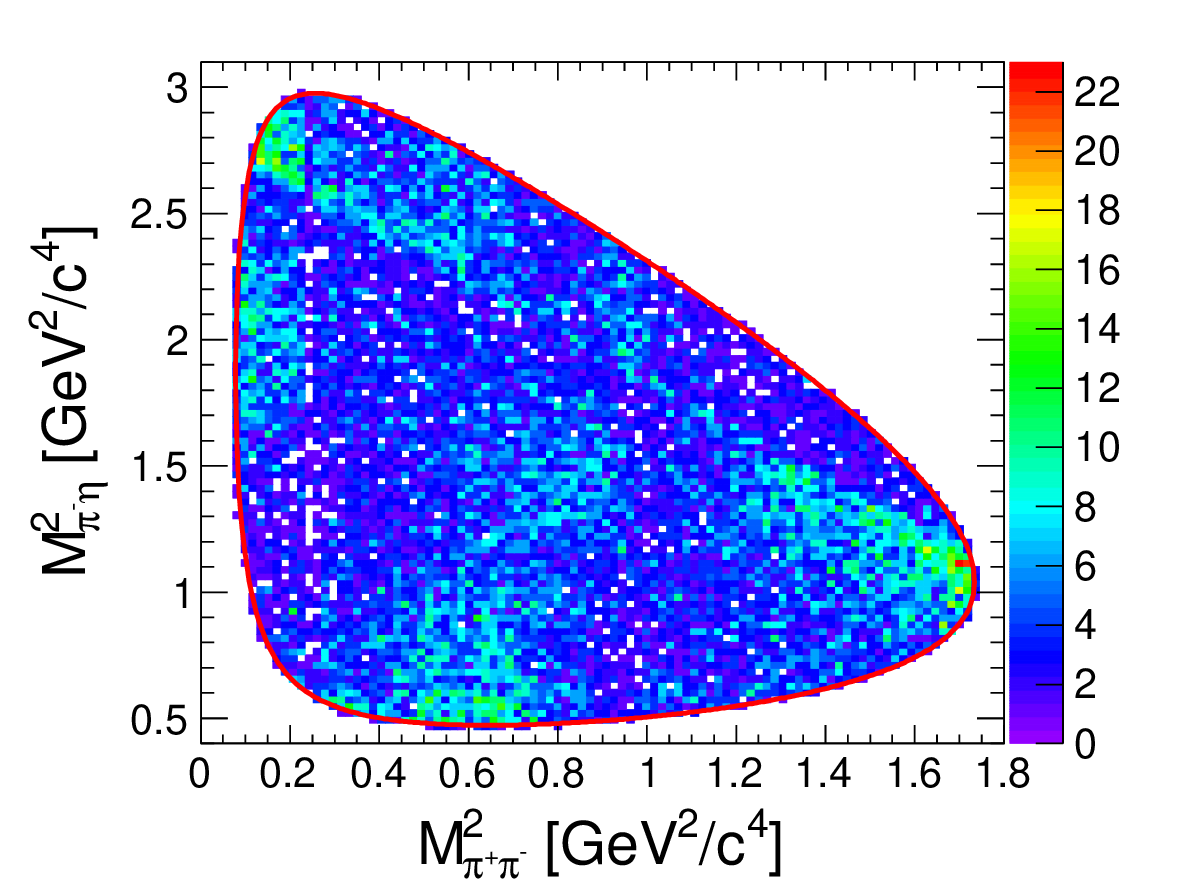}%
  \put(70,62){(b)}%
  \end{overpic}%
  \begin{overpic}[width=0.33\textwidth]{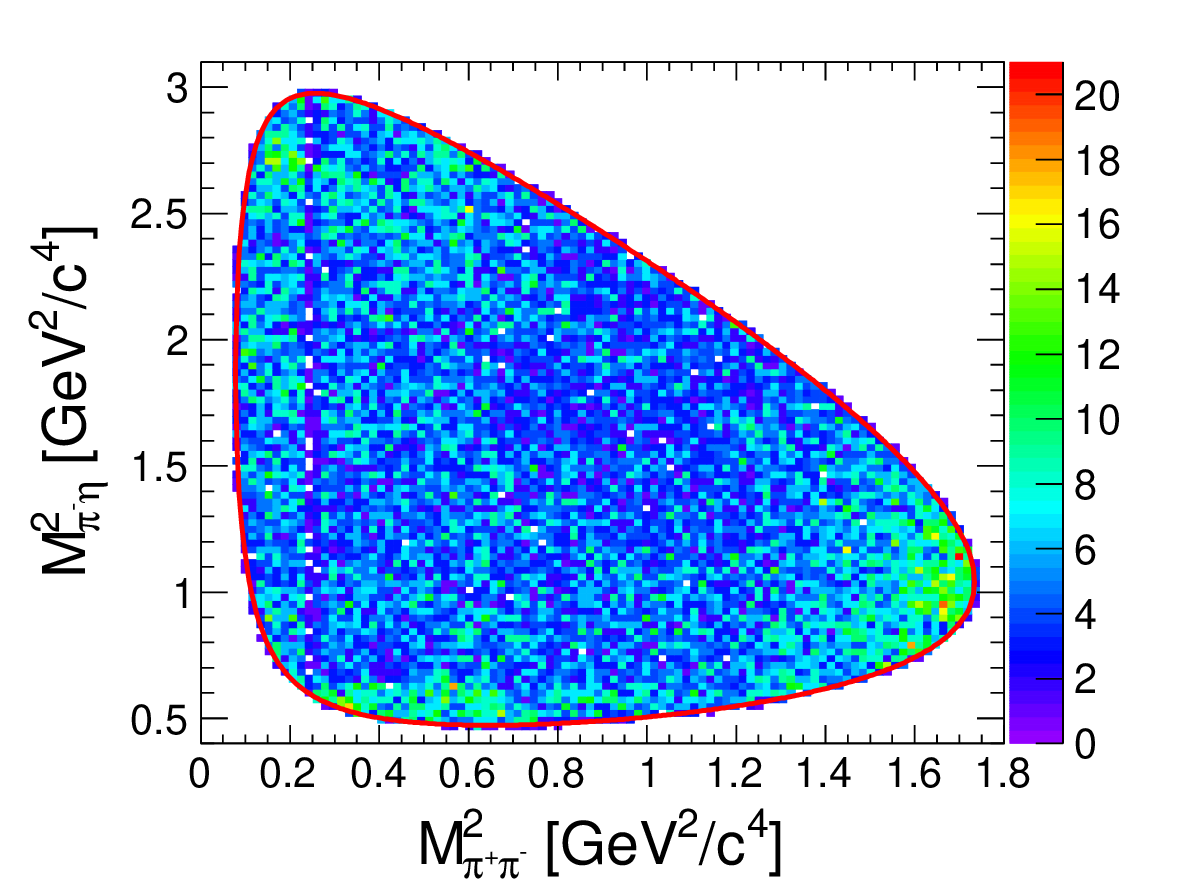}%
  \put(70,62){(c)}%
  \end{overpic}\\
  \begin{overpic}[width=0.33\textwidth]{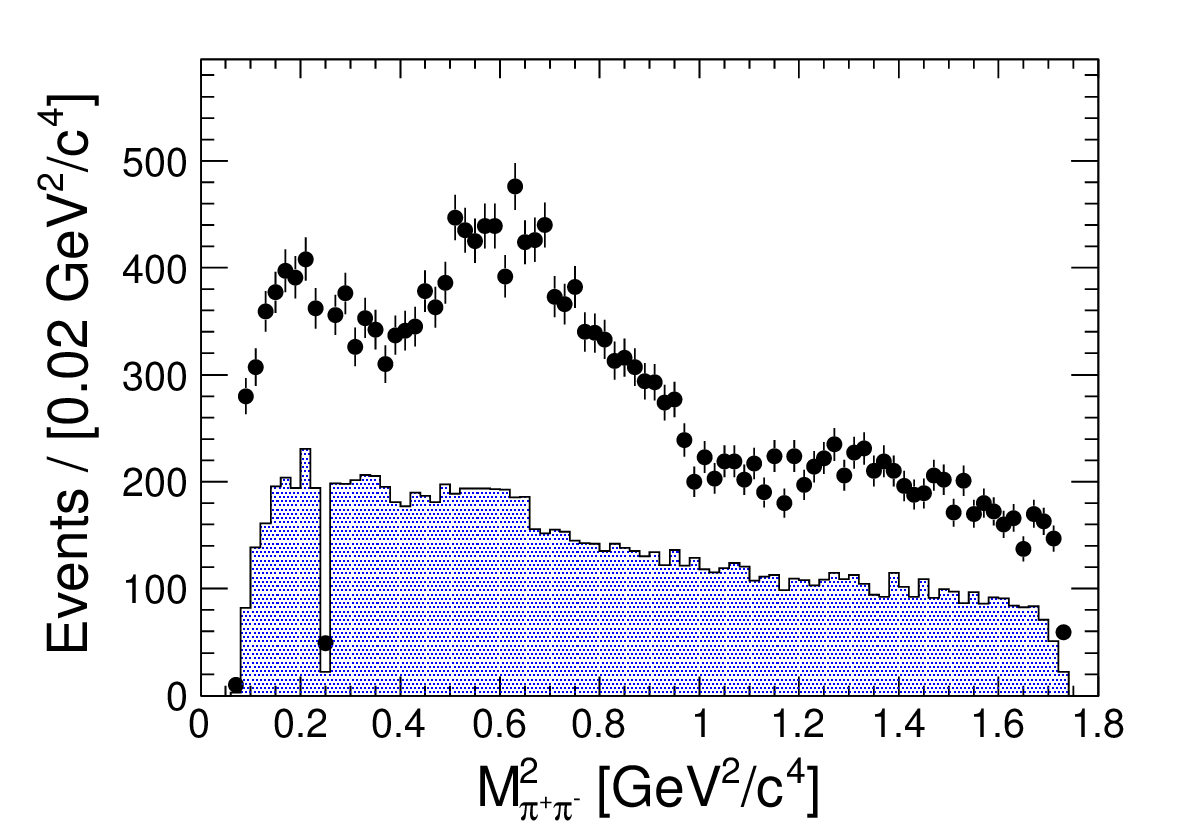}%
  \put(70,57){(d)}%
  \end{overpic}%
  \begin{overpic}[width=0.33\textwidth]{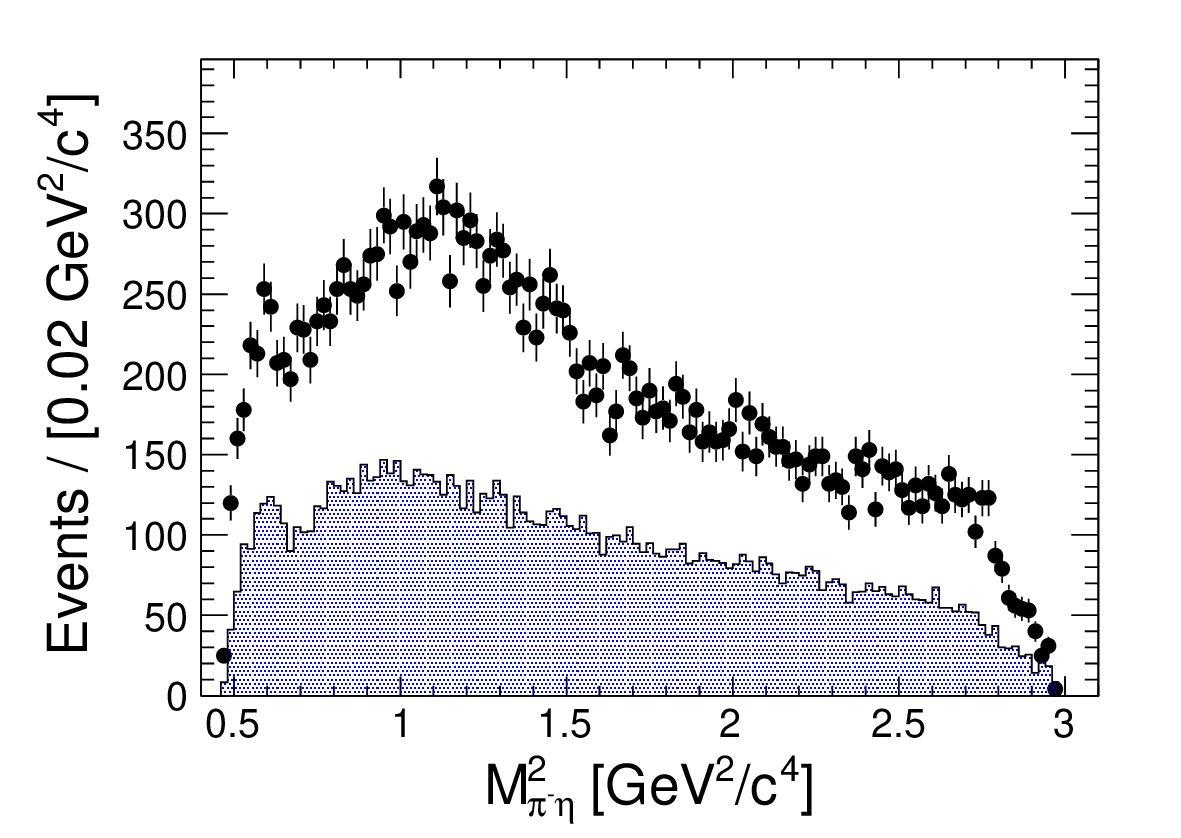}%
  \put(70,57){(e)}%
  \end{overpic}%
  \begin{overpic}[width=0.33\textwidth]{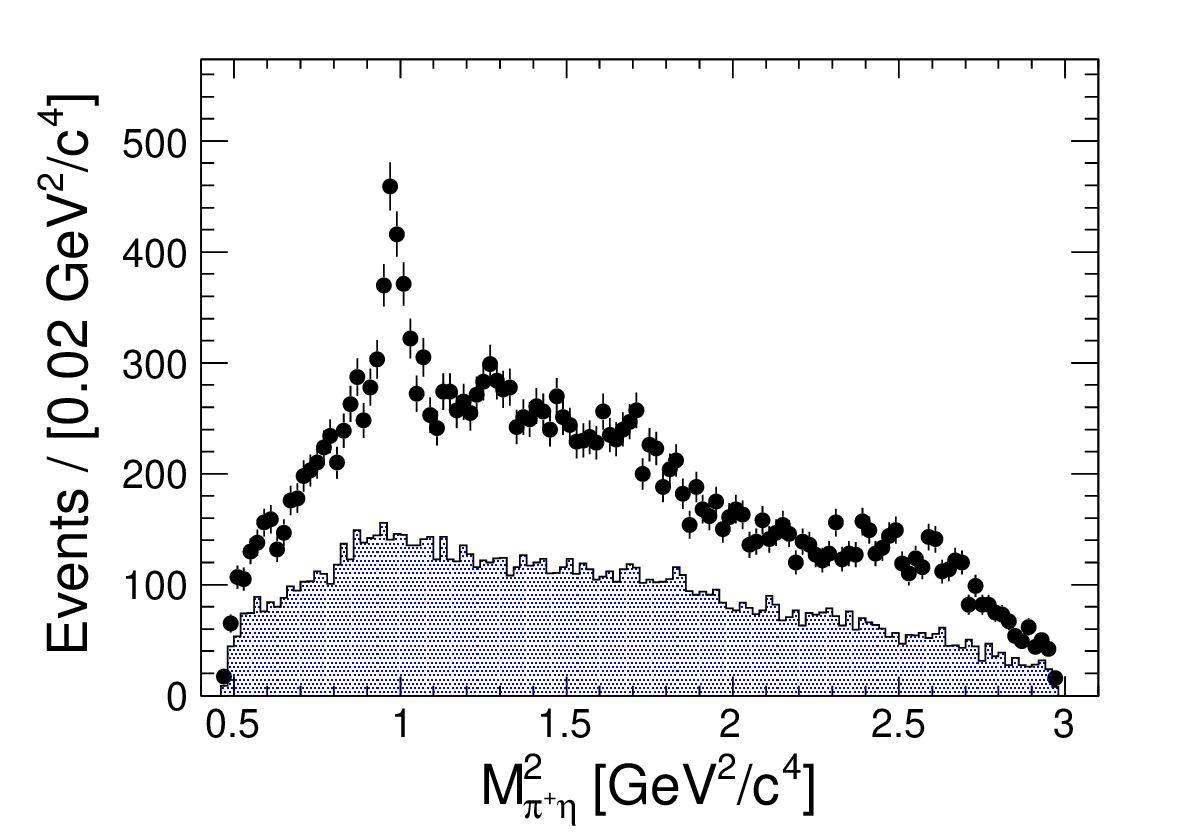}%
  \put(70,57){(f)}%
  \end{overpic}%
  \vskip-5pt
  \caption{For $\DzToPiPiEta$: (a) distribution of reconstruction efficiencies over the Dalitz plot, divided into $10\times 10$ bins of $M^{2}_{\pip\pim}$ vs $M^{2}_{\pim\eta}$. The red lines indicate the Dalitz plot boundaries. (b) Dalitz plot for events in the $Q$ signal region $|Q-5.86|<0.80$ MeV. (c) Dalitz plot for 
  events in the sideband region $2.5<|Q-5.86|<4.90~{\rm MeV}$, used to estimate the background shape. (d, e, f) Projections of Dalitz variables $M_{\pip\pim}^2$, $M_{\pim\eta}^2$, and $M_{\pip\eta}^2$, respectively. Points with error bars show events in the signal region; blue-filled histograms show the estimated background (see text). The dip in $M^2_{\pip\pim}$ near $0.25~{\rm GeV}^{2}/c^4$ is due to the $\KS$ veto.}
  \label{fig:unblindDzToPiPiEta}
  \end{centering}
\end{figure}

\begin{figure}[!hbtp]
  \begin{centering}%
  \begin{overpic}[width=0.33\textwidth]{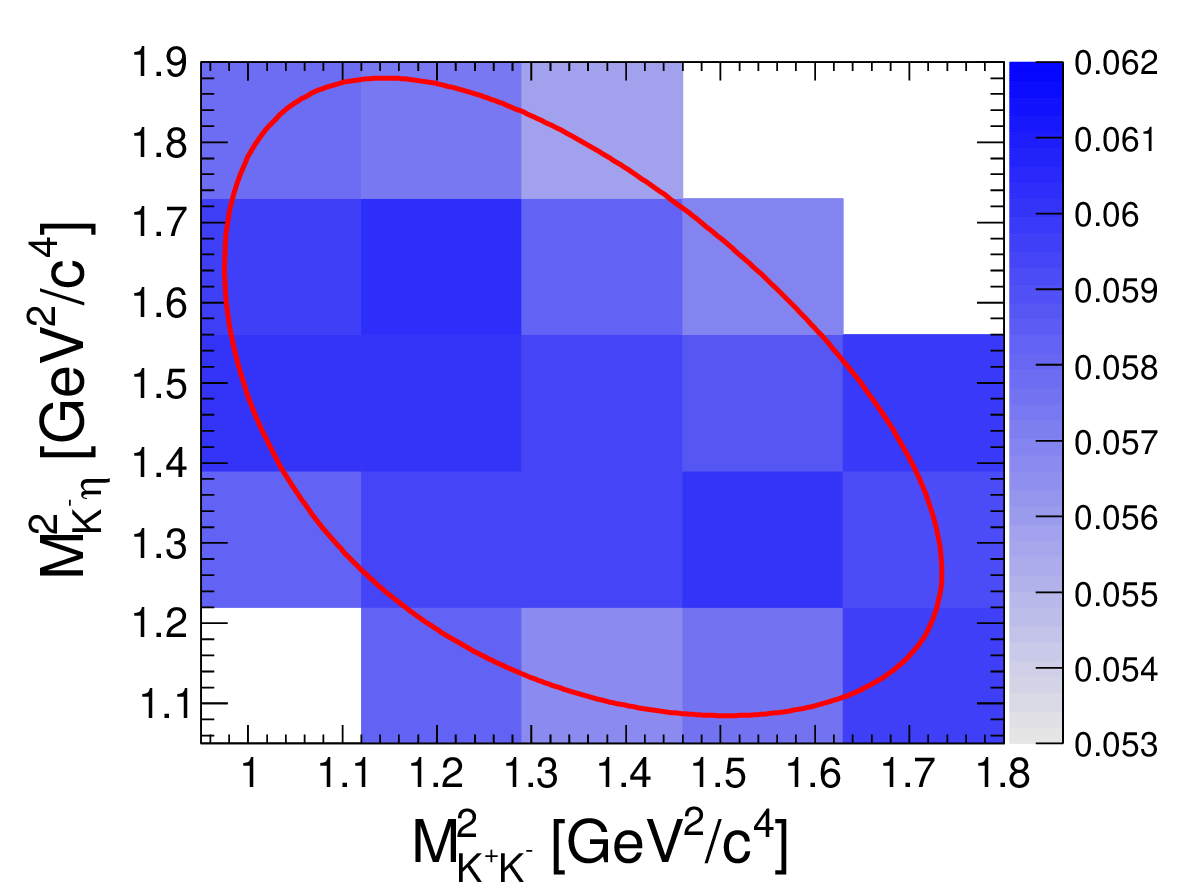}%
  \put(70,62){(a)}%
  \end{overpic}%
  \begin{overpic}[width=0.33\textwidth]{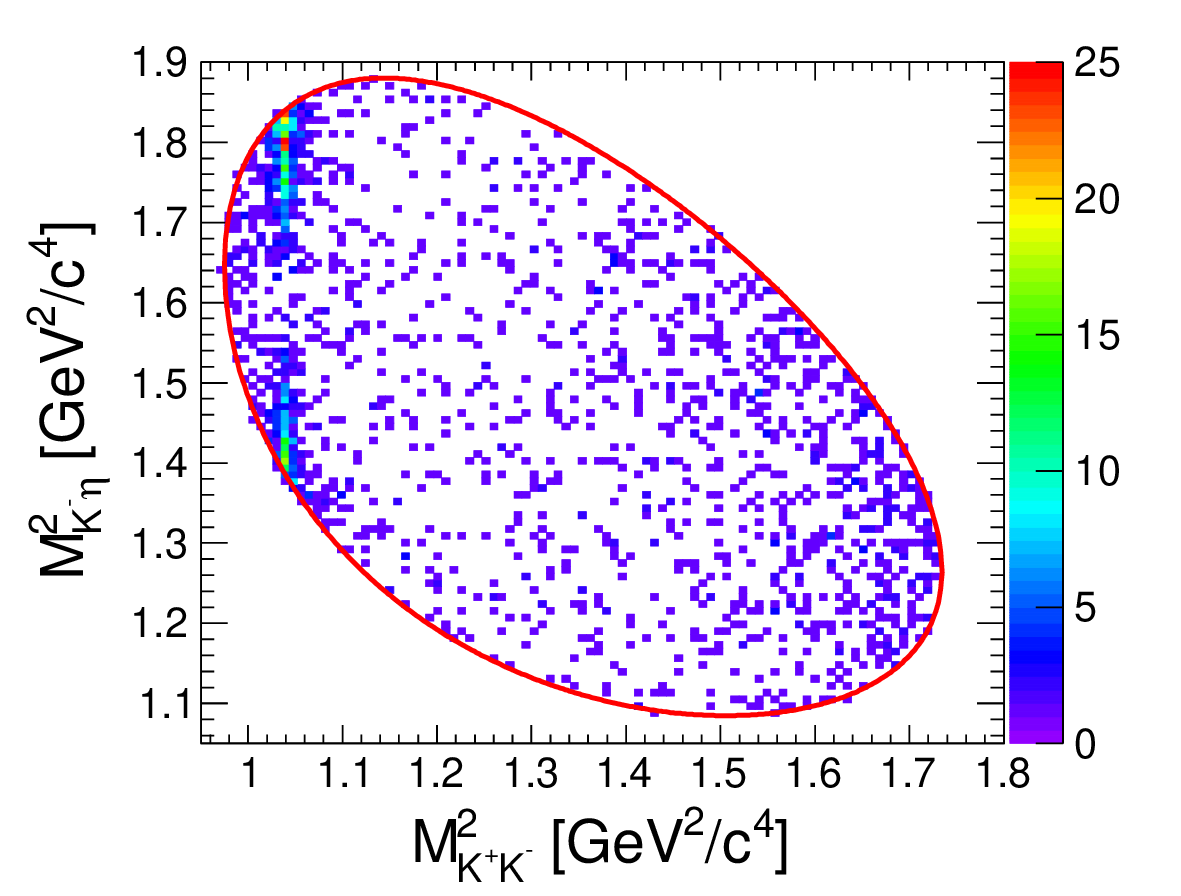}%
  \put(70,62){(b)}%
  \end{overpic}%
  \begin{overpic}[width=0.33\textwidth]{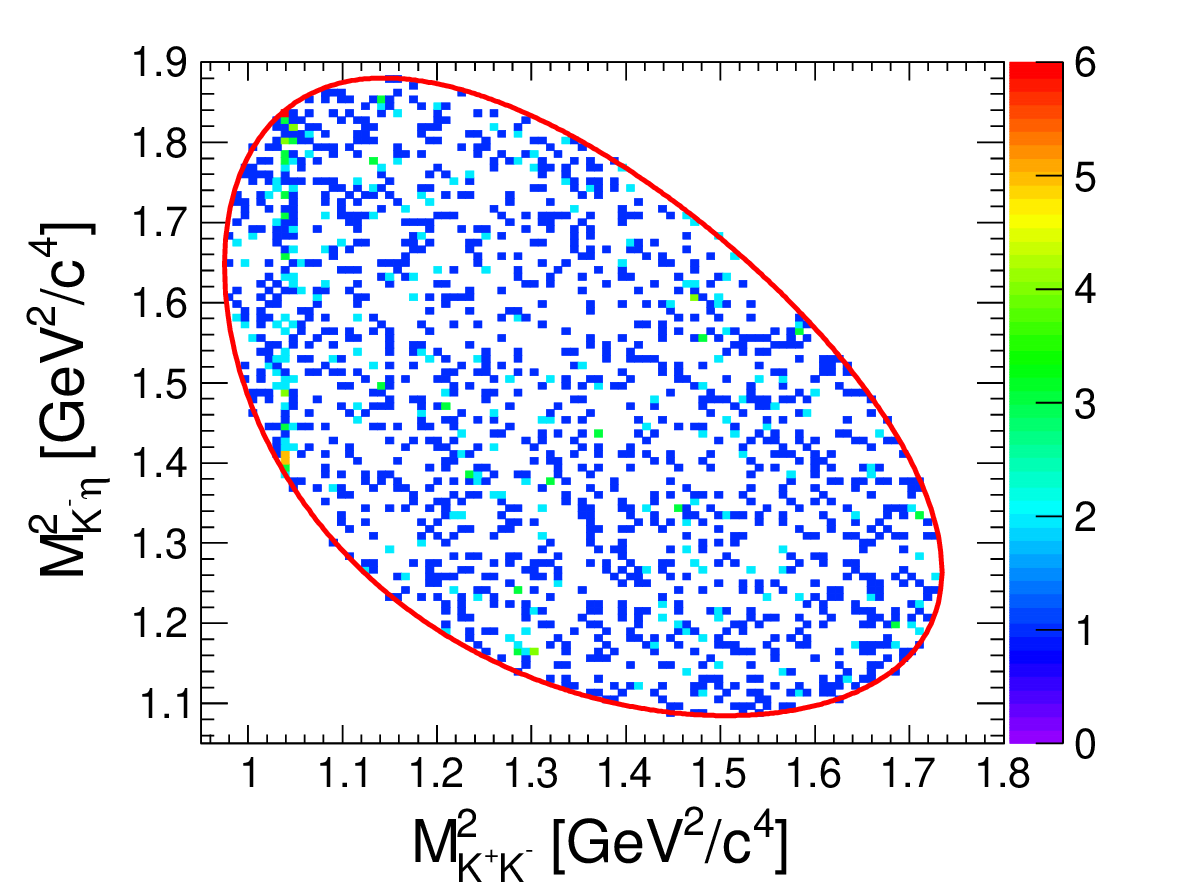}%
  \put(70,62){(c)}%
  \end{overpic}\\
  \begin{overpic}[width=0.33\textwidth]{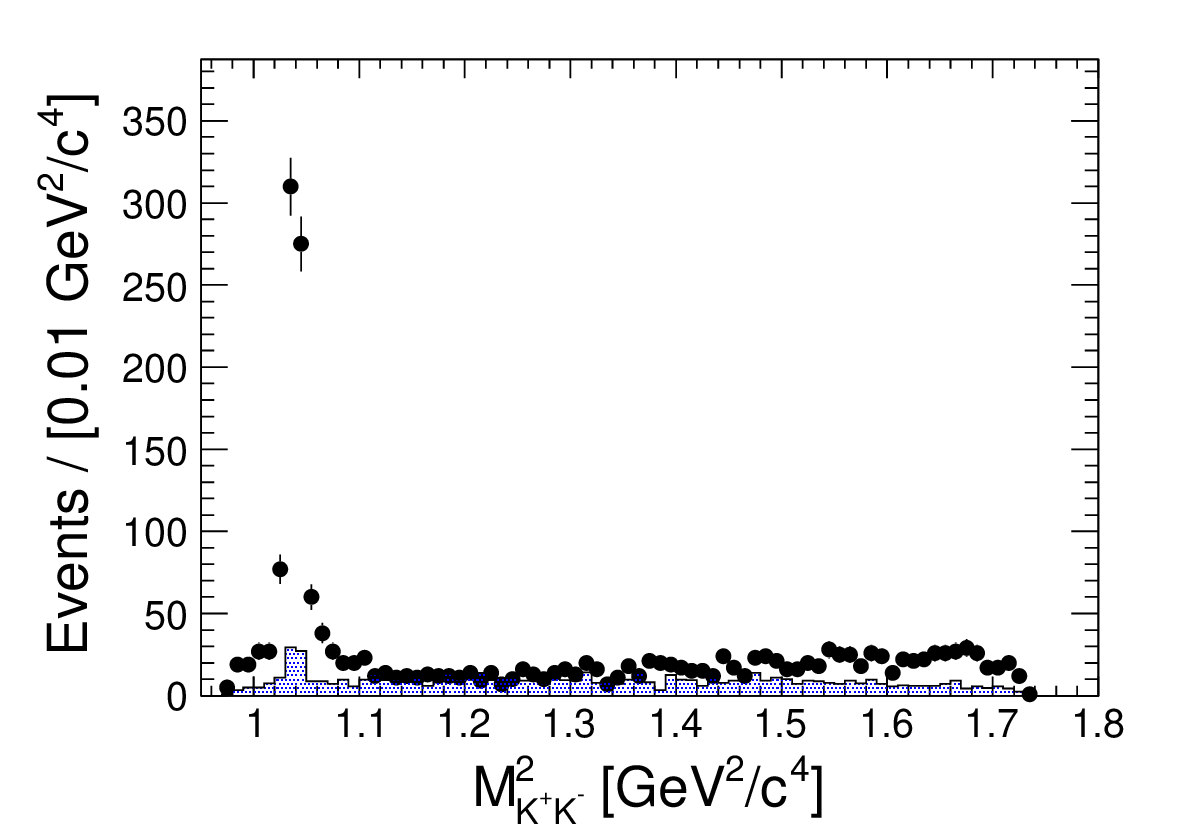}%
  \put(70,57){(d)}%
  \end{overpic}%
  \begin{overpic}[width=0.33\textwidth]{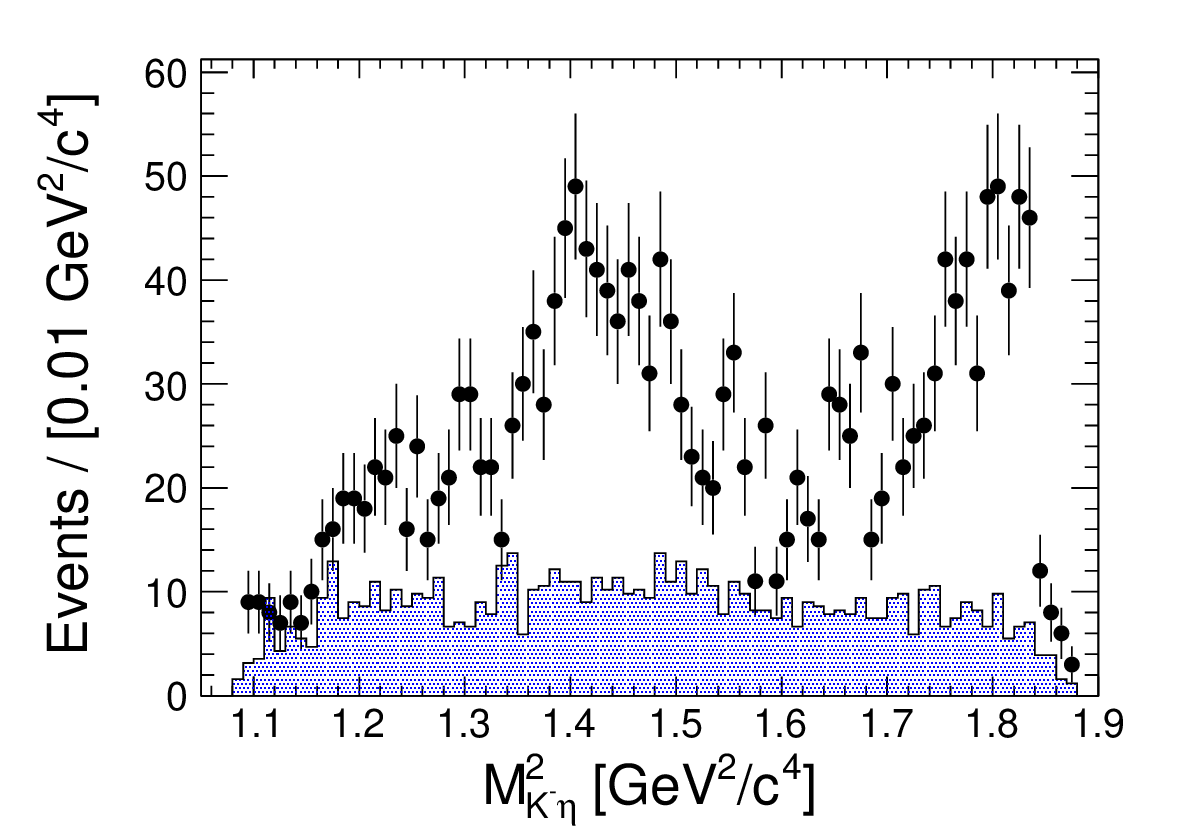}%
  \put(70,57){(e)}%
  \end{overpic}%
  \begin{overpic}[width=0.33\textwidth]{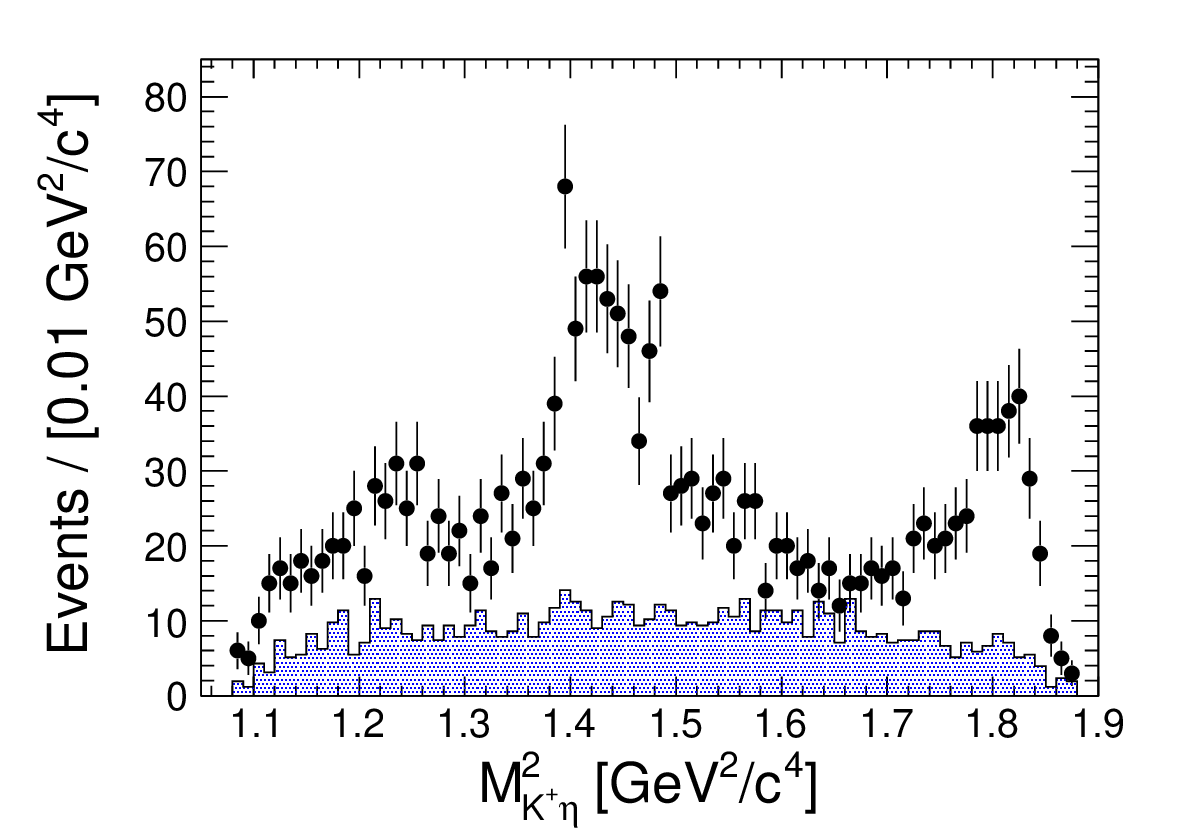}%
  \put(70,57){(f)}%
  \end{overpic}%
  \vskip-5pt
  \caption{For $\DzToKKEta$: 
  (a) distribution of reconstruction efficiencies over the Dalitz plot, divided into $5\times 5$ bins of $M^{2}_{\Kp\Km}$ vs $M^{2}_{\Km\eta}$. The red lines indicate the Dalitz plot boundaries. (b) Dalitz plot for events in the $Q$ signal region $|Q-5.86|<0.80$ MeV. (c) Dalitz plot for 
  events in the sideband region $2.5<|Q-5.86|<4.90$ MeV, used to estimate the background shape. (d, e, f) Projections of Dalitz variables $M_{\Kp\Km}^2$, $M_{\Km\eta}^2$, and $M_{\Kp\eta}^2$, respectively. Points with error bars show events in the signal region; blue-filled histograms show the estimated background (see text).
  }
  \label{fig:unblindDzToKKEta}
  \end{centering}
\end{figure}

To measure the branching fraction, we must divide these signal yields
by their reconstruction efficiencies. However, the reconstruction
efficiency for a decay can vary across the Dalitz plot of three-body phase space, and the Dalitz-plot distribution of $\DzToPiPiEta$ and
$\DzToKKEta$ decays has not been previously measured. Thus, 
to avoid systematic uncertainty due to the unknown 
Dalitz distribution (or decay model), 
we correct our signal yields for reconstruction
efficiencies as follows. We divide the Dalitz plot of the data into bins 
of $M^2(h^+h^-)$ and $M^2(h^-\eta)$, where $h=\pi$ or $K$, 
determine the reconstruction efficiency independently for each bin, 
and calculate the corrected signal yield via the formula
\begin{eqnarray}
N^{\rm cor} & =  & 
\sum_i \frac{N_i^{\text{tot}}-N^{\rm bkg}f_i^{\rm bkg}}
{\varepsilon_i}\,,  
\label{eqn:correctedYields}
\end{eqnarray}
where $i$ runs over all bins. 
The values of $M^2(h^+h^-)$ and $M^2(h^-\eta)$ 
are calculated subject to the constraint
$M(h^+h^-\eta) = m^{}_{D^0}$.
The formula (\ref{eqn:correctedYields}) has the following terms:
\begin{itemize}
\item $\varepsilon_i$ is the signal reconstruction efficiency for bin~$i$, as determined from a large sample of MC events. The resolutions in $Q$
of the MC samples are adjusted to match those of the data.
The efficiencies for $\DzToPiPiEta$ are plotted 
in figure~\ref{fig:unblindDzToPiPiEta}(a), and
those for $\DzToKKEta$ are plotted in figure~\ref{fig:unblindDzToKKEta}(a).
These efficiencies include a small ($\sim$\,2\%)
correction for $K^{\pm}$ and $\pi^{\pm}$ 
particle identification (PID) efficiencies, 
to account for small differences
observed between data and MC simulation. This correction is 
determined using a sample of $D^{*+}\to[\Dz\to\Km\pip]\pis$ decays.
\item $N_i^{\text{tot}}$ is the number of events in the $Q$ signal
region and the $i^{th}$ bin of the Dalitz plot. 
These yields are plotted in 
figure~\ref{fig:unblindDzToPiPiEta}(b) for
$\DzToPiPiEta$ and in 
figure~\ref{fig:unblindDzToKKEta}(b) for $\DzToKKEta$.
\item $N^{\rm bkg}$ is the total background yield in 
the $Q$ signal region, as obtained from fitting the 
$Q$ distribution (see figure~\ref{fig:unBlindQ}).
\item $f_i^{\rm bkg}$ is the fraction of background in the
$i^{\rm th}$-bin, with $\sum_i f_i=1$. These fractions are obtained
from the Dalitz plot distribution of events in the 
$Q$ sideband region $2.5~{\rm MeV}<|Q-5.86|<4.9~{\rm MeV}$.
The distribution of sideband events is shown in 
figure~\ref{fig:unblindDzToPiPiEta}(c) for
$\DzToPiPiEta$ and in 
figure~\ref{fig:unblindDzToKKEta}(c) for $\DzToKKEta$.
\end{itemize}
There are $10\times 10= 100$ bins in total for $\DzToPiPiEta$,
and $5\times 5 = 25$ bins total for $\DzToKKEta$. The final
corrected yields obtained using eq.~\eqref{eqn:correctedYields}~
are $N^{\rm cor} = (1.536\,^{+0.021}_{-0.020})\times 10^{5}$ 
for $\DzToPiPiEta$, and 
$N^{\rm cor} = (2.263\,^{+0.084}_{-0.077})\times 10^{4}$ 
for $\DzToKKEta$.

The branching fraction of a signal mode relative to that of
the normalization mode is determined from the ratio of 
their respective efficiency-corrected yields: 
\begin{eqnarray}
\frac{\mathcal{B}(D^0\to h^+h^-\eta)}{\mathcal{B}(\DzToKPiEta)} = 
\frac{N^{\rm cor}(D^0\to h^+h^-\eta)}{N^{\rm cor}(\DzToKPiEta)}\,, \label{eqn:eff}
\end{eqnarray}
where $h= K$ or $\pi$.
The efficiency-corrected yield for the normalization channel
$\DzToKPiEta$ is evaluated in a different manner than those of
the signal modes. As the Dalitz plot of 
$\DzToKPiEta$ decays has been measured with 
high statistics~\cite{bib:PRD102d012002}, we use the resulting
decay model to generate an MC sample, and use that sample
to evaluate the overall reconstruction efficiency.
The result, including the small PID efficiency correction, is 
$\varepsilon^{}_{K\pi\eta}=(6.870\pm 0.014)\%$.
Dividing the fitted yield for $\DzToKPiEta$ (see table~\ref{tabQ:yields_unblind2}) by this value gives 
$N^{\rm cor}(\DzToKPiEta)=(2.365\pm0.011)\times10^{6}$.

Inserting all efficiency-corrected yields into 
eq.~\eqref{eqn:eff}~gives the ratios of branching fractions 
\begin{eqnarray}
\frac{\mathcal{B}(\DzToPiPiEta)}{\mathcal{B}(\DzToKPiEta)} & = & 
[6.49 \pm 0.09\,({\rm stat}) \pm 0.12\,({\rm syst})]
\times10^{-2}\,,	\label{eqn:brrel_ppeta}	\\
\frac{\mathcal{B}(\DzToKKEta)}{\mathcal{B}(\DzToKPiEta)} & = & [9.57^{+0.36}_{-0.33}\,({\rm stat}) \pm 0.20\,({\rm syst})]
\times10^{-3}\,.  \label{eqn:brrel_KKeta}
\end{eqnarray}
The second error listed is the systematic uncertainty, which
is evaluated below (section~\ref{sec:sysBR}).
Multiplying both sides of eqs.~\eqref{eqn:brrel_ppeta}~and~\eqref{eqn:brrel_KKeta}~by the world average value 
$\mathcal{B}(\DzToKPiEta)=(1.88\pm 0.05)\%$~\cite{bib:PDG2020} gives
\begin{eqnarray}
\mathcal{B}(\DzToPiPiEta) & = & 
[1.22\pm 0.02\,({\rm stat}) \pm 0.02\,({\rm syst}) 
\pm 0.03\,(\mathcal{B}^{}_{\rm ref})]\times 10^{-3}\,, 
\label{eqn:brabs_ppeta}  \\
\mathcal{B}(\DzToKKEta) & = &  
[1.80\,^{+0.07}_{-0.06}\,({\rm stat}) \pm 0.04\,({\rm syst}) 
\pm 0.05\,(\mathcal{B}^{}_{\rm ref})] \times 10^{-4}\,,  
\label{eqn:brabs_KKeta}
\end{eqnarray}
where the third uncertainty listed is due to the branching 
fraction for the reference mode $\DzToKPiEta$. 
The result~\eqref{eqn:brabs_ppeta}~is consistent with the world average value $(1.17\pm0.07)\times10^{-3}$~\cite{bib:PDG2020}
but has improved precision. The result~\eqref{eqn:brabs_KKeta}~is the first such measurement.

The Dalitz plots and projections are shown in figure~\ref{fig:unblindDzToPiPiEta} for $\DzToPiPiEta$ and  in figure~\ref{fig:unblindDzToKKEta} for $\DzToKKEta$.
The background plotted is taken 
from the $Q$ sideband region, with the entries scaled 
to match the background yield in the signal region 
obtained from the $Q$ fit (figure~\ref{fig:unBlindQ}).
Several intermediate structures are clearly visible.
For $\DzToPiPiEta$ events, the $M^2_{\pip\pim}$ projection in figure~\ref{fig:unblindDzToPiPiEta}(d) shows 
the $D^0\ra \rho^0(770)\eta,\,\rho(770)\to\pip\pim$ decay process 
to be dominant. 
The $M^2(\pip\eta)$ distribution in figure~\ref{fig:unblindDzToPiPiEta}(f) shows a sharp peak near 
1.0~GeV$^{2}$/$c^{4}$, which indicates
$\Dz\ra a_0(980)^+\pim\!,\,a_0(980)^+\to\pip\eta$ decay. 
In contrast, 
the $M^2(\pim\eta)$ distribution in figure~\ref{fig:unblindDzToPiPiEta}(e) shows no indication 
of $\Dz\ra a_0(980)^-\pip\!,\,a_0(980)^-\to\pim\eta$.
This is unexpected, as the branching fraction for 
$\Dz\to a_0(980)^-\pip$ is predicted to be two 
orders of magnitude larger than that for 
$\Dz\to a_0(980)^+\pim$~\cite{bib:PRD67d034024}. 

For $\DzToKKEta$ events, the $M^2_{\Kp\Km}$ distribution shows 
the $D^0\ra \phi\eta,\,\phi\to\Kp\Km$ decay process to be dominant. 
However, a non-$\phi$ contribution is also visible. 
We thus measure $\mathcal{B}(\DzToKKEta)_{\phi{\rm -excluded}}$ by 
requiring $|M^{}_{\Kp\Km}\!- m^{}_{\phi}|>20$~MeV/$c^2$.
The signal yield is obtained as before by 
fitting the $Q$ distribution. The result is
$599\pm 45$ events in the signal region, 
as shown in figure~\ref{fig:unBlindQ}(d). 
The change in likelihood, with and without including a signal component in such $Q$ fitting, is $\Delta\ln L=214$.
As the number of degrees of freedom for the fit with no signal component is three less than that for the nominal fit (parameters $N_{\rm sig}$, $\mu$, and $\sigma_0$ are dropped), 
this value of $\Delta\ln L$ corresponds to a statistical significance for the signal of $20\sigma$.

We divide the signal yields obtained for bins of the Dalitz plot
by the efficiencies for these bins 
[see eq.~\eqref{eqn:correctedYields}]
to obtain
$N^{\rm cor}=12443\,^{+1071}_{-893}$ (for $\DzToKKEta$ with the $\phi$ excluded).
Thus
\begin{eqnarray}
\frac{\mathcal{B}(\DzToKKEta)_{\phi{\rm -excluded}}}
{\mathcal{B}(\DzToKPiEta)} & = & 
[5.26\,^{+0.45}_{-0.38}\,(\rm{stat})\pm 0.11\,({\rm syst})]\times10^{-3}\,. \label{eqn:BRnonphiEta}
\end{eqnarray}
The second error listed is the systematic uncertainty, which
is evaluated below (section~\ref{sec:sysBR}).
Multiplying each side of eq.~\eqref{eqn:BRnonphiEta}~by 
$\mathcal{B}(\Dz\to\Km\pip\eta)=(1.88\pm0.05)\%$~\cite{bib:PDG2020} gives the branching fraction for $\DzToKKEta$ with 
the $\phi$ component excluded by requiring $|M^{}_{KK}\!- m^{}_{\phi}|>20~{\rm MeV}/c^2$:  
\begin{eqnarray}
\hskip-0.30in
\mathcal{B}(\DzToKKEta)_{\phi{\rm -excluded}}
& = & [0.99\,^{+0.08}_{-0.07}\,({\rm stat})\pm 0.02\,({\rm syst})\pm  0.03\,(\mathcal{B}_{\rm ref})]\times 10^{-4}.\label{eqn:phi-vetoBR}
\end{eqnarray}
This result is somewhat higher (but more precise) 
than a similar measurement by BESIII, 
$(0.59\pm 0.19)\times 10^{-4}$~\cite{bib:PRL124d241803}.

\subsection{\boldmath{Measurement of $\mathcal{B}(D^0\to\phi\eta)$}}\label{sec:BF2}
As shown in figure~\ref{fig:unblindDzToKKEta}, 
the decay $\DzToKKEta$ is dominated by the 
Cabibbo- and color-suppressed 
decay $\DzToPhiEta,\,\phi\to K^+ K^-$.
We thus measure the branching fraction for
$\DzToPhiEta$ by performing a two-dimensional fit to
the $M_{KK}$ and $Q$ distributions of
$\DzToKKEta$ events. The fitted region 
is $M_{KK}<1.08$ GeV/$c^2$ and $Q<15$ MeV. 
In this region, signal decays and background are
straightforward to identify: the
non-$\phi$ $\DzToKKEta$ component peaks in $Q$ but not in $M_{KK}$,
whereas combinatorial background containing $\phi\to K^+K^-$ decays peak in $M_{KK}$ but not in $Q$.
The signal PDF for $M_{KK}$ is taken to be the sum of a Gaussian 
and two asymmetric Gaussians, with a common mean for the $M(\phi)$ peak. The PDF for $Q$ is taken to be a bifurcated 
Student's t-function: 
\begin{eqnarray}
\mathcal{P}_{\rm sig}(M_{KK}) 	& = & f_2 [ f_1 G(\mu_m, \sigma_{m0}) + (1-f_1) G_{\rm asym}(\mu_m, r_1\sigma_{m0}, \delta_2) ] \nonumber \\
    & & + (1-f_2) G_{\rm asym}(\mu_m, r_2r_1\sigma_{m0}, \delta_3)\,, \label{eqn:MphiQ2D_sig0} \\
\mathcal{P}_{\rm sig}(Q) & = &  S_{\rm bif}(\mu_q, \sigma_0, \delta_0, n_l, n_h)\,,   \label{eqn:MphiQ2D_sig1} \\ 
\mathcal{P}_{\rm sig}^{D^0\to\phi\eta}(M_{KK},Q) & = & \mathcal{P}_{\rm sig}(M_{KK})\cdot \mathcal{P}_{\rm sig}(Q)\,. \label{eqn:MphiQ2D_sig}
\end{eqnarray}
Most of the signal shape parameters are fixed to MC values;
the parameters $\mu_m,\,\sigma_{m0}$ for $M_{KK}$ and $\mu_q,\,\sigma_0$ 
for $Q$ are floated to account for possible differences in 
resolution between data and MC simulation.

The non-$\phi$ $\DzToKKEta$ component includes several 
processes such as $\Dz\to a_0(980)\eta$, 
$\Dz\to K_0^{*}(1430)^{\pm}K^{\mp}$, and non-resonant 
$\DzToKKEta$ decays. 
We parameterize this component in $M_{KK}$ by a
threshold function, and in $Q$ by the same PDF as that used
for the signal:
\begin{eqnarray}
\mathcal{P}_{\rm peak}(M_{KK}) 	& = & \sqrt{M_{KK}-m_0} \cdot e^{-\beta_{\rm peak} (M_{KK}-m_0)}\,,  \\
\mathcal{P}_{\rm peak}(Q) 		& = &   \mathcal{P}_{\rm sig}(Q)\,,  \\ 
\mathcal{P}_{\rm peak}(M_{KK},Q) & = & \mathcal{P}_{\rm peak}(M_{KK})\cdot \mathcal{P}_{\rm peak}(Q)\,,  \label{eqn:MphiQ2D_QpeakBkg}
\end{eqnarray}
where the threshold value $m_0=2m_K=0.987354$ GeV/$c^2$.
We consider possible interference between the non-$\phi$ component and the $\Dz\to\phi\eta$ signal as a 
systematic uncertainty.

For combinatorial background, the $Q$ distribution 
is parameterized with a threshold function. The $M_{KK}$
distribution has two parts: (1) a $\phi$-peak, which is taken
to be the same as that of signal decay, and (2) a threshold function. 
These PDFs take the forms
\begin{eqnarray}
\mathcal{P}_{\rm bkg}(M_{KK}) 	& = & f_{\phi} \mathcal{P}_{\rm sig}(M_{KK}) + (1-f_{\phi})  (M_{KK}-m_0)^{\alpha_m} e^{-\beta_m (M_{KK}-m_0)}\,,  \\
\mathcal{P}_{\rm bkg}(Q) 		& = &   Q^{\alpha_q} e^{-\beta_q Q}\,,   \\ 
\mathcal{P}_{\rm bkg}(M_{KK},Q) & = & \mathcal{P}_{\rm bkg}(M_{KK})\cdot \mathcal{P}_{\rm bkg}(Q)\,.  \label{eqn:MphiQ2D_bkg}
\end{eqnarray}
The relative fraction $f_{\phi}$ and the $Q$ threshold parameters are floated; all other parameters are fixed to MC values.

\begin{figure*}[!hbtp]
  \begin{centering}%
  \begin{overpic}[width=0.85\textwidth]{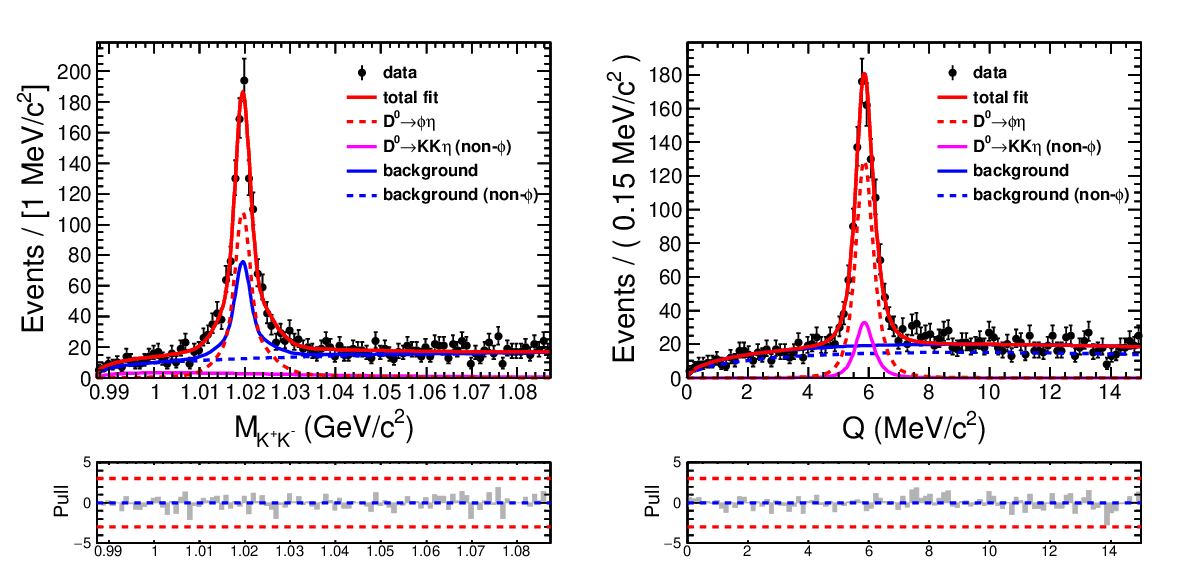}%
    \put(10,40){(a)}%
    \put(60,40){(b)}%
  \end{overpic}\\
  \begin{overpic}[width=0.85\textwidth]{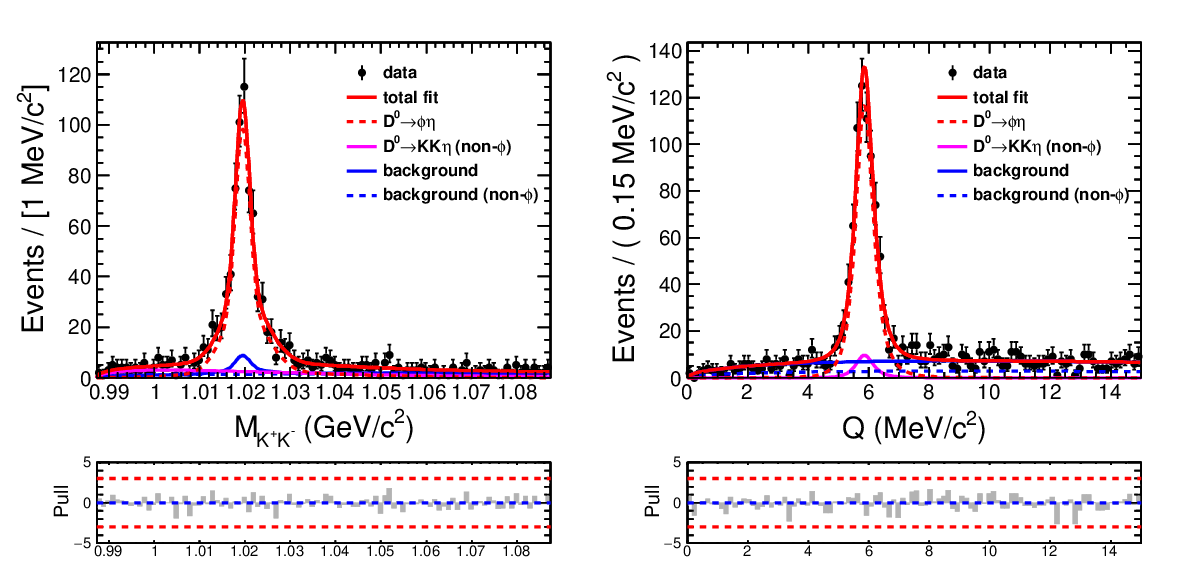}%
    \put(10,40){(c)}%
    \put(60,40){(d)}%
  \end{overpic}%
   \vskip-5pt
 \caption{\label{fig:MphiQ_unblind}Projections of $\Kp\Km$ invariant mass distributions in $Q$ (a) fit region and (c) signal region ($|Q-5.86|<0.8$ MeV) and $Q$ distributions in $M_{\Kp\Km}$ (b) fit region and (d) signal region ($|M_{\Kp\Km}-m_{\phi}|<10$ MeV/$c^{2}$) from the $M_{KK}$-$Q$ two-dimensional fit. Points with error bars are the data. The red solid line is total fit result. The red dashed curves are signal of $\DzToPhiEta$, and the magenta solid curves show the $Q$-peaking background from non-$\phi$ component in $\DzToKKEta$. The blue dash line is the non-$\phi$ component of total non-$Q$-peaking background (blue line).} 
  \end{centering}
\end{figure*}

The results of the two-dimensional likelihood fit are shown in figure~\ref{fig:MphiQ_unblind}. We obtain a signal yield 
$N_{\rm sig}=728\pm 36$ in the full fitted region, and 
$N_{\rm sig}=600\pm29$ in the signal region 
$|Q-5.86|<0.8$ MeV and $|M_{\Kp\Km}-m_{\phi}|<10$ MeV/$c^{2}$.
The difference in likelihood, with and without 
including a signal component, is 
$\Delta\ln\mathcal{L}=464.8$. 
As the number of degrees of freedom for the fit with no signal component is one less than that for the nominal fit (parameter $N_{\rm sig}$ is dropped), this value of $\Delta\ln L$ 
corresponds to a statistical significance for $\DzToPhiEta$ of $31\sigma$. 

We evaluate the signal reconstruction efficiency using
a large MC sample of $\Dz\to\phi\eta$ decays. We obtain,
for events in the $M_{KK}$-$Q$ signal region, 
an efficiency $\varepsilon=(5.262\pm 0.021)\%$. 
Thus, $N^{\rm cor}(\Dz\to\phi\eta,\,\phi\to\Kp\Km) =(1.140\pm 0.055)\times10^{4}$,
and
\begin{eqnarray}
\frac{\mathcal{B}(D^0\to\phi\eta,\,\phi\to\Kp\Km)}
{\mathcal{B}(\Dz\to\Km\pip\eta)} 
 = [4.82\pm 0.23\,({\rm stat})\pm 0.16\,({\rm syst})]
 \times 10^{-3}\,.
\end{eqnarray}
The second error listed is the systematic uncertainty, which
is evaluated below (section~\ref{sec:sysBR}).
Multiplying each side by the world average value
$\mathcal{B}(\Dz\to\Km\pip\eta)=(1.88\pm 0.05)\%$~\cite{bib:PDG2020}
and dividing by $\mathcal{B}(\phi\to\Kp\Km)=(49.2\pm 0.5)\%$~\cite{bib:PDG2020}, we obtain 
\begin{eqnarray}
\mathcal{B}(D^0\to\phi\eta)=
[1.84\pm 0.09\,({\rm stat})\pm 0.06\,({\rm syst})\pm 0.05\,(\mathcal{B}_{\rm ref})]\times 10^{-4}\,,
\end{eqnarray}
where the systematic uncertainty includes the small uncertainty on $\mathcal{B}(\phi\to\Kp\Km)$.
This result is consistent with, but notably more precise than, the current world average of 
$(1.8\pm0.5)\times10^{-4}$~\cite{bib:PDG2020}.
It is also consistent with theoretical predictions~\cite{bib:PRD100d093002,bib:PRD89d054006}.
As a consistency check, 
we calculate the branching fraction 
of the non-$\phi$ $\DzToKKEta$ component by subtracting 
the $\DzToPhiEta$ branching fraction from the total 
$\DzToKKEta$ result:
\begin{eqnarray}
\mathcal{B}(\DzToKKEta)-\mathcal{B}(\DzToPhiEta,\phi\to\Kp\Km) = (0.90\pm 0.08)\times 10^{-4}\,,
\end{eqnarray}
which is very close to our measurement of
$\mathcal{B}(\DzToKKEta)_{\phi{\rm -excluded}}$ in eq.~\eqref{eqn:phi-vetoBR}.

\subsection{Systematic uncertainties}
\label{sec:sysBR}
The sources of systematic uncertainty in measuring the
branching fractions are listed in table~\ref{tab:sysBF}.
These uncertainties are evaluated as follows.
\begin{table*}[!htbp]
\begin{center}
\begin{tabular}{|l|c|c|c|} \hline 
Systematic sources 	& $\frac{\mathcal{B}(\DzToPiPiEta)}{\mathcal{B}(\DzToKPiEta)}$  
								& $\frac{\mathcal{B}(\DzToKKEta)}{\mathcal{B}(\DzToKPiEta)}$  
											& $\frac{\mathcal{B}(D^0\to(\phi\to\Kp\Km)\eta)}{\mathcal{B}(\Dz\to\Km\pip\eta)}$ \\ \hline
PID efficiency correction		
				& 	1.8\%		&   1.9\%	&	1.9\%	\\
Signal PDF		& 	0.3\%		& 	 0.5\%	&	0.9\%	\\
Background PDF 	& 	0.0\%		& 	 0.0\%	& 	0.1\%	\\
Mass resolution	 calibration
				&       0.1\%		&	0.3\%	&	0.0\%	\\
Yield correction with efficiency map &  	0.3\%		& 	0.7\% 	&	--	\\
MC statistics		&      	0.3\%		&	0.4\%	&	0.4\%	\\
$K_S^0$ veto		& 	0.1\%		& 	--		&	-- 		\\ 
Interference in $M_{KK}$	&	--			&	--		& 	 2.5\%		\\\hline
Total syst.	error		&	1.9\%		&	2.1\%	&	3.3\%	\\ \hline 
\end{tabular}
\caption{Systematic uncertainties (fractional) for the branching ratio measurements.}
\label{tab:sysBF}
\end{center}  
\end{table*} 

\begin{itemize}
\item A correction for PID efficiency is applied to 
$K^{\pm}$ and $\pi^{\pm}$ 
tracks, to account for a difference in efficiency
between data and MC simulation. 
The correction depends on track momentum and is small; the
uncertainty on the correction is even smaller, in the
range~(0.90-0.97)\%. 
When evaluating this uncertainty for a ratio of branching
fractions, we conservatively assume the efficiency corrections
for $\Kp$ and $\pip$ tracks (which appear separately in 
numerator and denominator, or vice-versa) are anticorrelated.

\item The uncertainty due to the 
parameters fixed in the fit for the signal yield 
is evaluated as follows.
We sample these parameters simultaneously from
Gaussian distributions,
accounting for their correlations, 
and re-fit for 
the signal yield. The procedure is repeated 1000 times and these yields are plotted. The ratio of the
root-mean-square (RMS) to the mean value of the resulting distribution 
of signal yields is taken as the
systematic uncertainty due to the fixed parameters.
The Gaussian distributions from which the parameters
are sampled have mean values equal to the fixed values of the parameters, and widths equal 
to their respective uncertainties.

\item 
For background PDFs, all parameters are floated in the fits
except for those describing the amount of background and its 
shape for $\DzToKPiEta$, which are taken from MC simulation. 
We evaluate this uncertainty due to these fixed parameters as done above,
by simultaneously sampling these parameters from 
Gaussian distributions having mean values equal to
the fixed values and widths equal to their respective 
uncertainties. The
RMS of the resulting distribution of $\DzToKPiEta$ 
yields is taken as the systematic uncertainty due 
to the peaking background.

\item We correct for differences in mass resolutions (including $M$, $Q$ and $M_{KK}$)
between data and MC when calculating reconstruction efficiencies (in eq.~\eqref{eqn:correctedYields}~for $\DzToPiPiEta$ and $\DzToKKEta$, as well as for $\DzToPhiEta$ and the reference mode). We take the systematic uncertainty of 
this procedure to be the difference in the ratio of efficiency-corrected signal yields to that of the reference mode obtained both with and without this resolution correction.

\item The efficiency for $\DzToPiPiEta$ and $\DzToKKEta$ decays is
evaluated in bins of the Dalitz plot; see
eq.~\eqref{eqn:correctedYields}. This efficiency has 
uncertainty arising from the number of bins used, from the 
efficiency values $\varepsilon^{}_i$ for the various bins,
and from the bin-by-bin background subtraction. 
\begin{itemize}
\item For the first uncertainty, we vary the numbers of 
bins used, and the corresponding change in the efficiency 
is taken as the systematic uncertainty. 
For $\DzToPiPiEta$ decays, our nominal result uses 
$10\times10$ bins; thus we also try $8\times8$ and 
$12\times12$ bins.
For $\DzToKKEta$ decays, our nominal result uses 
$5\times5$ bins; thus we also try $3\times3$ and 
$7\times7$ bins. We obtain 0.25\% for $\DzToPiPiEta$,
and 0.50\% for $\DzToKKEta$.

\item to evaluate the effect of uncertainties
in $\varepsilon^{}_i$, we sample the $\varepsilon^{}_i$
from Gaussian distributions having mean values equal 
to the nominal values, and widths equal to their 
uncertainties. For each
sampling, we re-calculate the yield $N^{\rm cor}$
using eq.~\eqref{eqn:correctedYields} and plot the 
result. The RMS of this distribution is taken
as the systematic uncertainty:
$0.21\%$ for $\DzToPiPiEta$, and $0.43\%$ for $\DzToKKEta$.

\item the background subtraction procedure depends on the
distribution of background over the Dalitz plot. We take
this distribution from a data $Q$ sideband region. To evaluate 
the uncertainty due to this Dalitz distribution, we shift 
the $Q$ sideband region used by $\pm0.4$ MeV and repeat 
the procedure. The change is assigned as the systematic uncertainty: $0.02\%$ for $\DzToPiPiEta$, and $0.03\%$ for $\DzToKKEta$.

\end{itemize}

\item The efficiency for $\DzToKPiEta$
is evaluated from MC simulation using a Dalitz decay model. 
The uncertainty due to this
model is evaluated by varying the model and 
re-calculating~$\varepsilon$. Specifically, our 
nominal model uses eight intermediate resonances as 
measured in ref.~\cite{bib:PRD102d012002}; as an alternative,
we include all thirteen intermediate resonances
listed in ref.~\cite{bib:PRD102d012002}. The resulting
change in our reconstruction efficiency is very small ($\delta\varepsilon/\varepsilon<0.01\%$).
As a cross check, we calculate $N^{\rm cor}$ for
$\DzToKPiEta$ using eq.~\eqref{eqn:correctedYields}; 
the result is 
$(2.369\pm 0.007)\times 10^{6}$, which is almost
identical with our nominal result 
(the difference is much smaller than the uncertainty).

\item There are small uncertainties in the reconstruction efficiencies
$\varepsilon$, which are evaluated from MC simulation, due to the finite statistics of the MC samples used.

\item  There is an uncertainty arising from the $\KS$ veto required for
$\DzToPiPiEta$ decays. We evaluate this by changing the veto
region from $m_{\KS}\pm 10$~MeV/$c^2$ ($\sim 3\sigma$) to 
$m_{\KS}\pm 15$~MeV/$c^2$ ($\sim5\sigma$); the resulting
change in the $\DzToPiPiEta$ signal yield is taken as 
the systematic uncertainty.

\item We consider possible interference between the 
$\DzToPhiEta$ amplitude and that of non-$\phi$
$\DzToKKEta$. Such interference could alter the $M_{KK}$
distribution used to fit for the $\DzToPhiEta$ yield.
We evaluate this effect by introducing a relative 
phase $\theta$ between the two amplitudes, which 
modifies the PDF to be
	\begin{eqnarray}
	\hskip-0.20in
	\mathcal{P}^{}_{\rm total}(M_{KK}, Q) & = & \left| A_{\phi}(M_{KK}) + r e^{i\cdot(\theta+k\cdot \pi)} \sqrt{ F_{{\rm non-}\phi}(M_{KK}) } \right|^2 \times F_{\rm sig}(Q)\,.
	\end{eqnarray} 
In this expression, $A_{\phi}$ is a relativistic Breit-Wigner 
function, $F_{{\rm non-}\phi}$ is the shape function
used in the nominal fit to model non-$\phi$ decays, 
and $k$ is a factor that adds
a phase shift of $\pi$ depending on whether the cosine
of the $K^+$ helicity angle ($\theta^{}_h$) is positive 
or negative, i.e., $k=0$ for $\cos\theta^{}_h<0$, and $k=1$
otherwise~\cite{bib:PRD71d092003}.
The helicity angle $\theta^{}_h$ is defined as the angle 
between the momentum of the $K^+$ and the 
$\eta$
in the $K^+K^-$ rest frame.
After fitting with this PDF, we calculate the 
$\DzToPhiEta$ yield as the product of the total yield
obtained and the fraction $f^{}_\phi$ given by
\begin{eqnarray}
	f_{\phi} & = & \int \left| A_{\phi} \right|^2 dM_{KK}/\int \left| A_{\phi} + r e^{i\cdot(\theta+k\cdot \pi)} \sqrt{ F_{\rm{non-}\phi}} \right|^2 dM_{KK}\,.
	\end{eqnarray}
The result is that the $\DzToPhiEta$ yield decreases by 2.5\%,
and thus we assign this value as the systematic uncertainty due 
to possible interference.

\end{itemize}
The total systematic uncertainty is obtained by adding 
in quadrature all the above contributions. The results
are listed in table~\ref{tab:sysBF}.

\section{\boldmath{Measurement of $\CP$ asymmetries}}
\label{sec:CPasym}

\subsection{\boldmath{Measurement of $\Acp(\DzToPiPiEta)$ and $\Acp(\DzToKKEta)$}}
To measure the $\CP$ asymmetries, we divide the sample for 
each channel into $\Dz$ and $\Dzb$ decays, where the flavor 
of the $\Dz$ or $\Dzb$ is tagged by the charge of the 
$\pi^\pm_s$ from the $D^{*+}\to\Dz\pi^+_s$ or 
$D^{*-}\to\Dzb\pi^-_s$ decay. 
To correct for an asymmetry 
in $\pi^\pm_s$ reconstruction efficiencies, we
weight events according to the $\pi^\pm_s$ efficiency
mapping of ref.~\cite{bib:PRL112d211601}.

\begin{figure*}[!hbtp]
  \begin{center}%
  \begin{overpic}[width=0.45\textwidth]{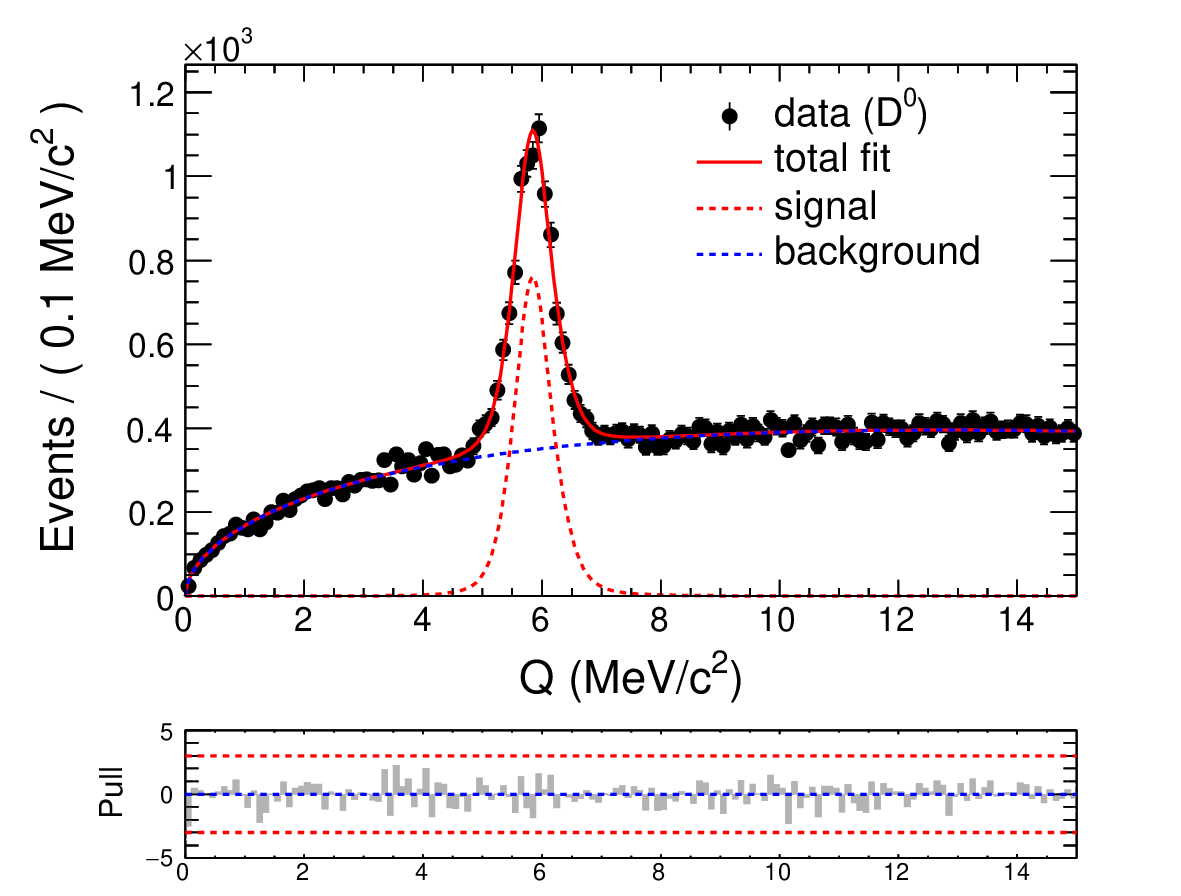}%
  \put(25,60){(a)}
  \end{overpic}%
  \begin{overpic}[width=0.45\textwidth]{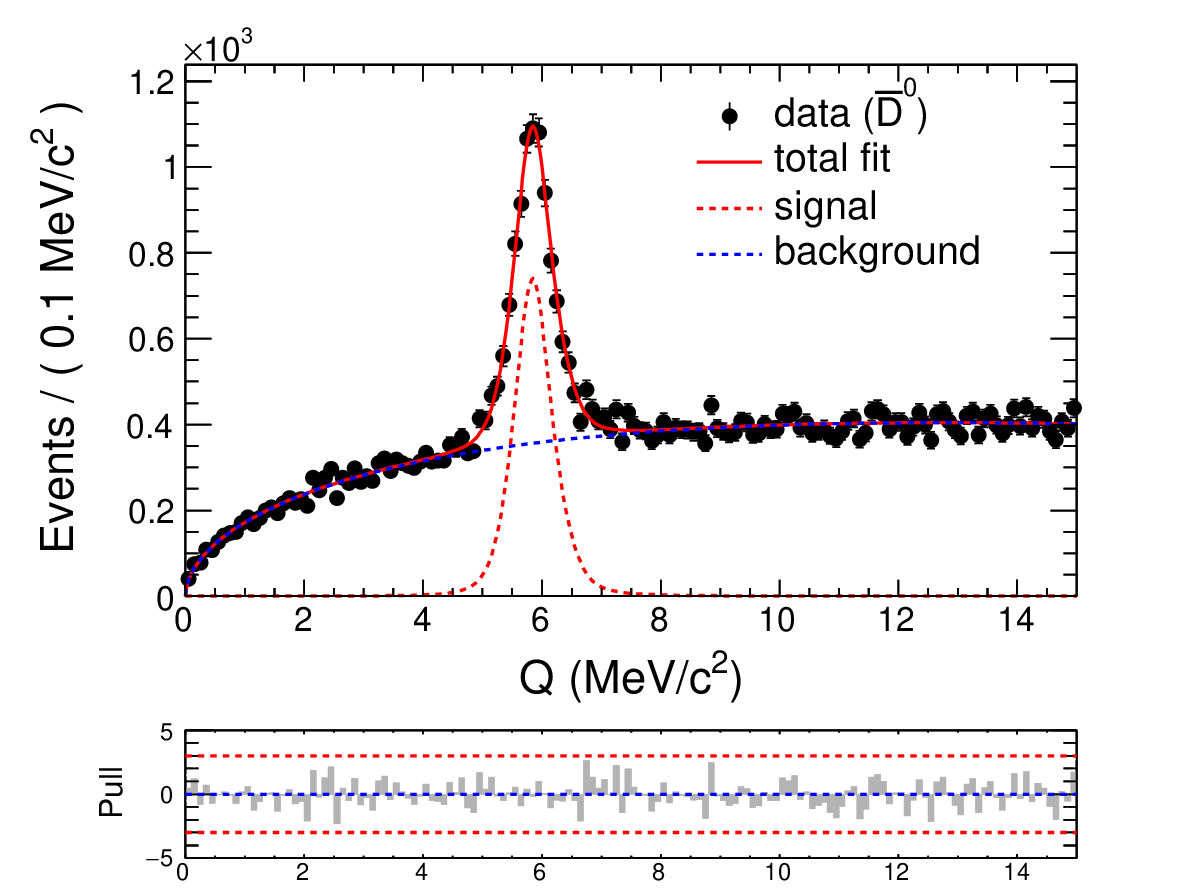}%
  \put(25,60){(b)}
  \end{overpic}
  \begin{overpic}[width=0.45\textwidth]{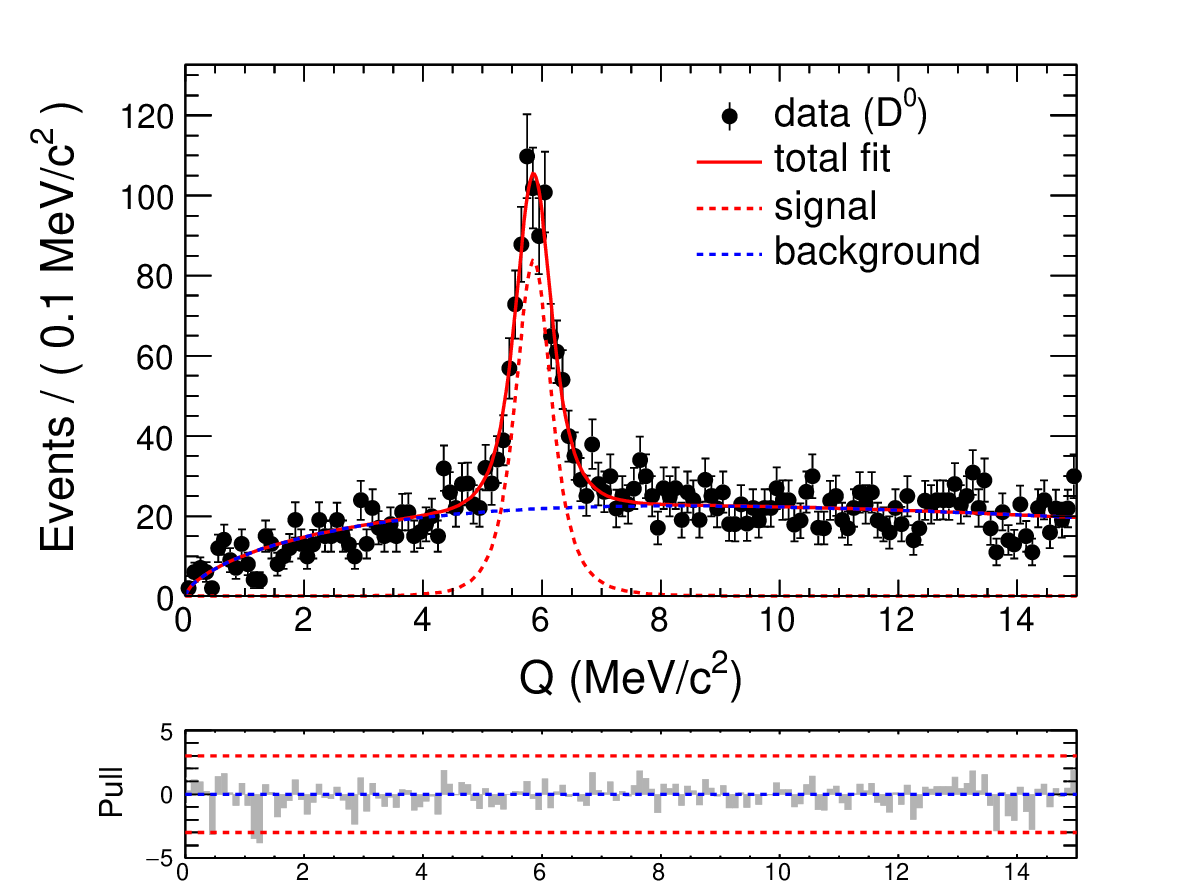}%
  \put(25,60){(c)}
  \end{overpic}%
  \begin{overpic}[width=0.45\textwidth]{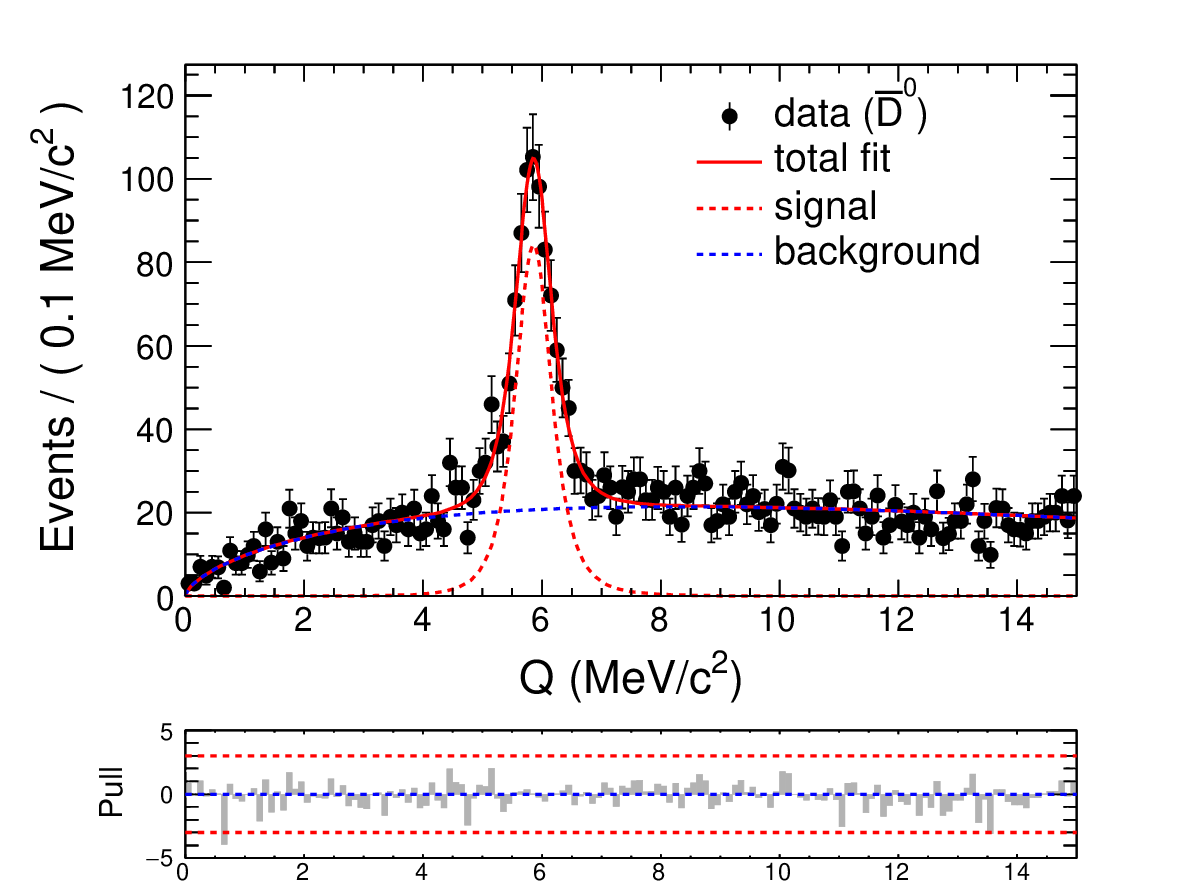}%
  \put(25,60){(d)}
  \end{overpic}  
  \caption{
  Simultaneous fit for $\DzToPiPiEta$ (a) and $\Dzb\to\pip\pim\eta$ (b) candidates; and $\DzToKKEta$ (c) and $\Dzb\to\Kp\Km\eta$ (d) candidates.}
  \label{fig:unblindAcpDzTohhEta}
  \end{center}
\end{figure*}

We simultaneously fit the $Q$ distributions
of these weighted samples for $\Dz$ and $\Dzb$ decays 
with parameters fixed in the same way as done for the branching fraction measurements.
The parameters $N^{}_{\rm sig}$ and $A_{\rm corr}$ are fitted,
where the $\Dz$ and $\Dzb$ signal yields are given by
$N_{\rm sig}(\Dz,\Dzb) = (N_{\rm sig}/2) \cdot 
(1\pm A_{\rm corr})$.
The fit results are shown in 
figure~\ref{fig:unblindAcpDzTohhEta}. We obtain
$N_{\rm sig}=12975\pm 198$ and $A_{\rm corr}=(1.44 \pm 1.24)\%$
for $\DzToPiPiEta$ decays, and 
$N_{\rm sig}=1482\pm 60$ and $A_{\rm corr}=(-0.25 \pm 2.96)\%$ 
for $\DzToKKEta$ decays.

These values for $A_{\rm corr}$ include the 
forward-backward asymmetry $A^{}_{\rm FB}$.
We correct for $A^{}_{\rm FB}$ by calculating
$A_{\rm corr}$ in eight bins of $\cos\theta^*$,
where $\theta^*$ is the polar angle of the $D^{*+}$ with
respect to the $+z$ axis in the $e^+e^-$ CM frame. 
The bins used are
$[-1.0,\,-0.6]$, 
$[-0.6,\,-0.4]$, 
$[-0.4,\,-0.2]$, 
$[-0.2,\,0.0]$, 
$[0.0,\,0.2]$, 
$[0.2,\,0.4]$, 
$[0.4,\,0.6]$, and 
$[0.6,\,1.0]$.
The asymmetries
$A^{}_{\CP}$ and $A^{}_{\rm FB}$ are then 
extracted via eqs.~\eqref{eqn:Acp1}~and~\eqref{eqn:Acp2}.
The resulting four values of $A^{}_{\CP}$ and $A^{}_{\rm FB}$ 
are plotted in 
figures~\ref{fig:Acp_cosTheta}(a,\,d) for $\DzToPiPiEta$ and in figures~\ref{fig:Acp_cosTheta}(b,\,e) for $\DzToKKEta$.
Fitting the $A^{}_{\CP}$ values to constants yields the
final results
\begin{eqnarray}
A_{\CP}(\DzToPiPiEta) & = & 
[0.9\pm 1.2\,({\rm stat})\pm 0.5\,({\rm syst})]\%\,, \\
A_{\CP}(\DzToKKEta) & = & 
[-1.4\pm 3.3\,({\rm stat})\pm 1.1\,({\rm syst})]\%\,.
\end{eqnarray} 
The second error listed is the systematic uncertainty, 
which is evaluated below (section~\ref{sec:systAcp}).
The first result is a factor of four more precise than 
a recent measurement by BESIII~\cite{bib:PRD101d052009},
while the latter result is the first such measurement. 
The $A^{}_{\rm FB}$ values plotted in 
figures~\ref{fig:Acp_cosTheta}(d-e) decrease with $\cos\theta^*$ and are consistent with (somewhat lower in $\DzToPiPiEta$ than) the
leading-order prediction~\cite{bib:ZPC30d124} at $\sqrt{s}=10.6$~GeV of 
$A_{\rm FB}^{c\bar{c}}=
-0.029\cdot\cos\theta^{*}/(1+\cos^2\theta^{*})$, at the current level of statistics.

\subsection{\boldmath{Measurement of $\Acp(\DzToPhiEta)$}}
\begin{figure*}[!hbtp]
  \begin{center}%
  \begin{overpic}[width=0.45\textwidth]{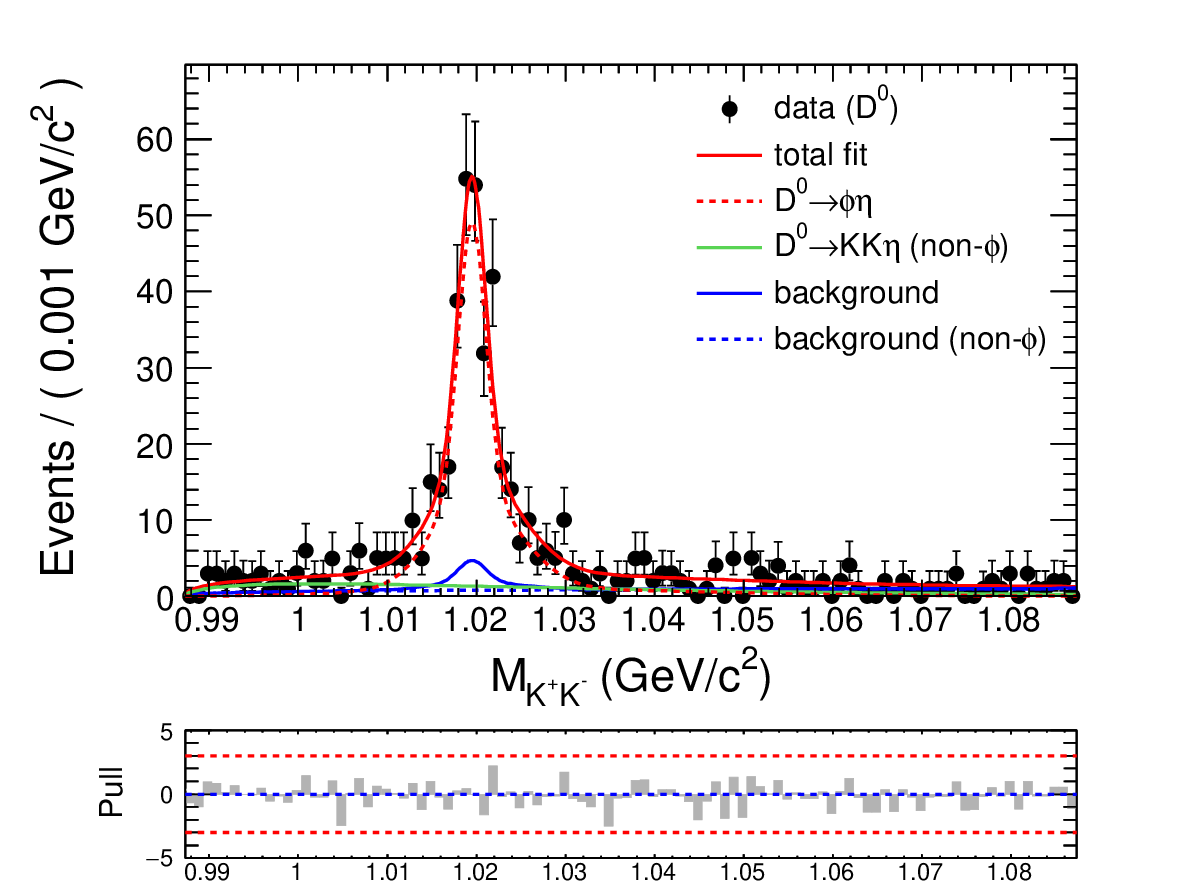}%
  \put(25,60){(a)}
  \end{overpic}%
  \begin{overpic}[width=0.45\textwidth]{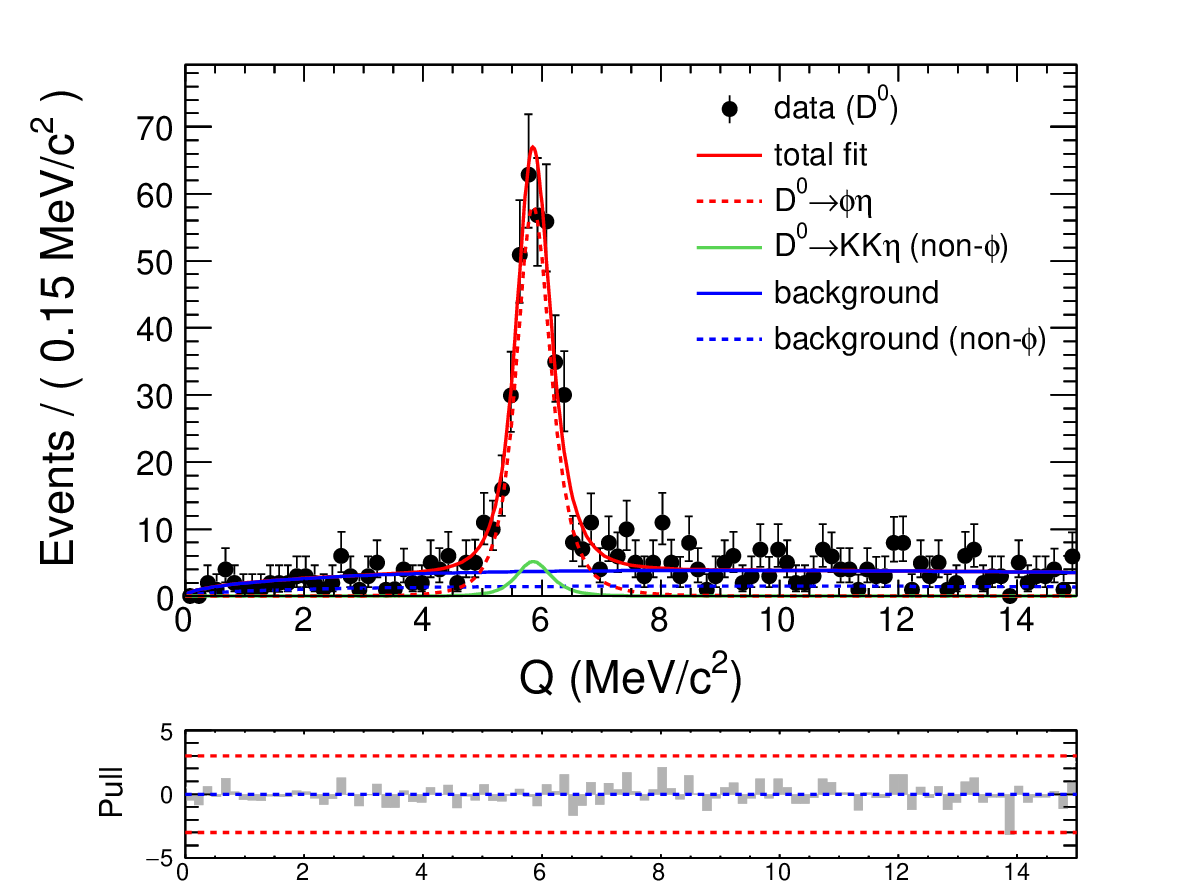}%
  \put(25,60){(b)}
  \end{overpic}\\
  \begin{overpic}[width=0.45\textwidth]{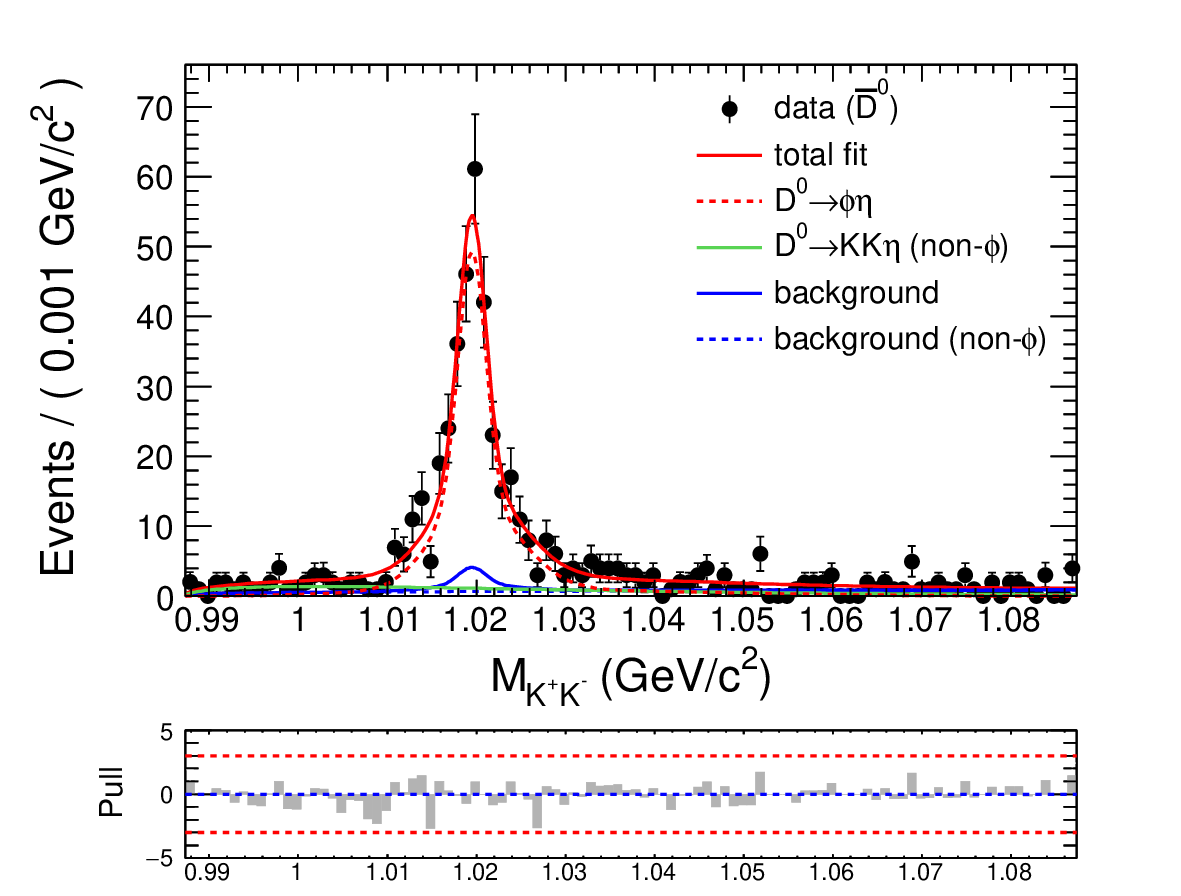}%
  \put(25,60){(c)}
  \end{overpic}%
  \begin{overpic}[width=0.45\textwidth]{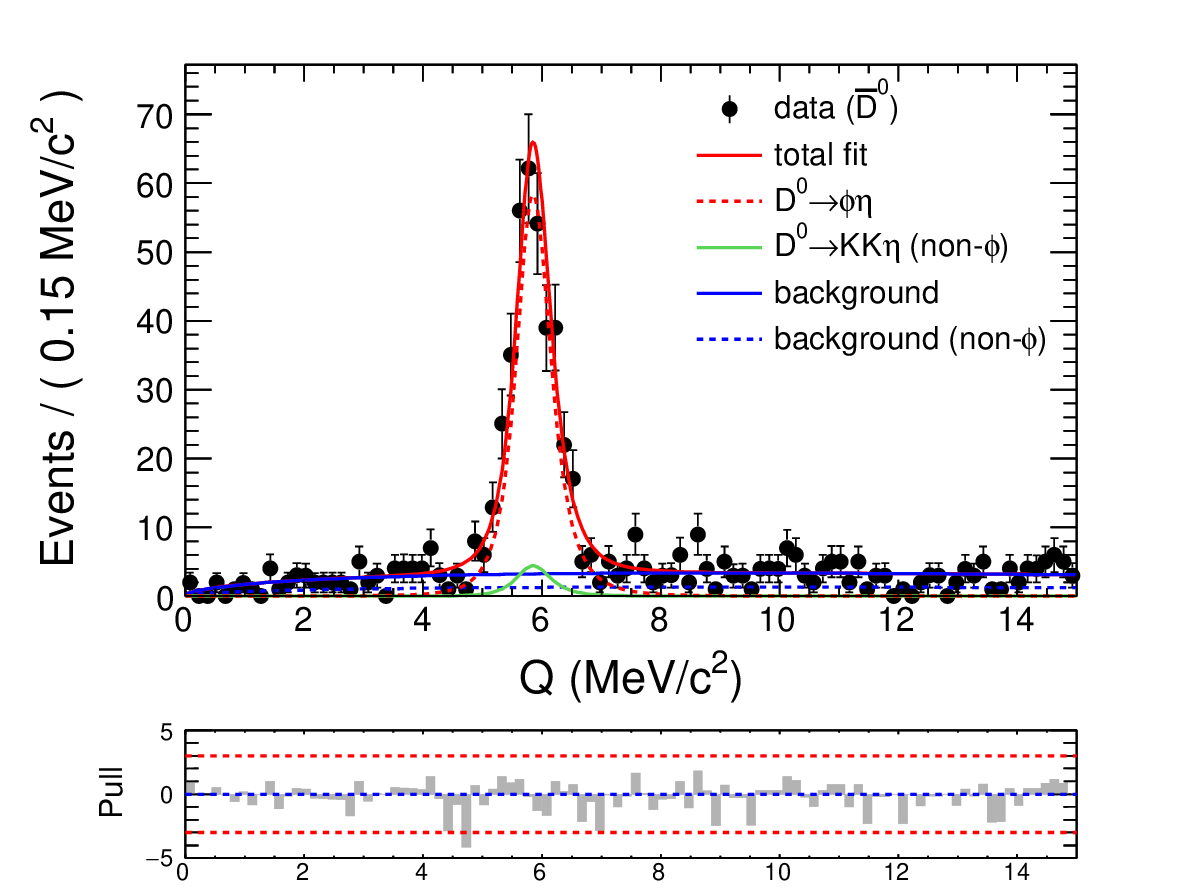}%
  \put(25,60){(d)}
  \end{overpic}%
  \caption{
  Projections in $M_{K^+K^-}$ (left) and $Q$ (right) of the two-dimensional $(M_{K^+K^-}, Q)$ fit, for $\DzToPhiEta$ (top) and $\Dzb\to\phi\eta$ (bottom). In both cases, the $M_{K^+K^-}$ projection corresponds 
 to the $Q$ signal region ($|Q-5.86|<0.8$ MeV), and the $Q$ projection corresponds to the
  $M_{K^+K^-}$ signal region ($|M_{K^+K^-}-m_{\phi}|<0.01$ GeV/$c^2$). 
} 
  \label{fig:unblindAcpDzToPhiEta}
  \end{center}
\end{figure*}

\begin{figure}[!hbtp]
  \begin{center}%
  \begin{overpic}[width=0.333\textwidth]{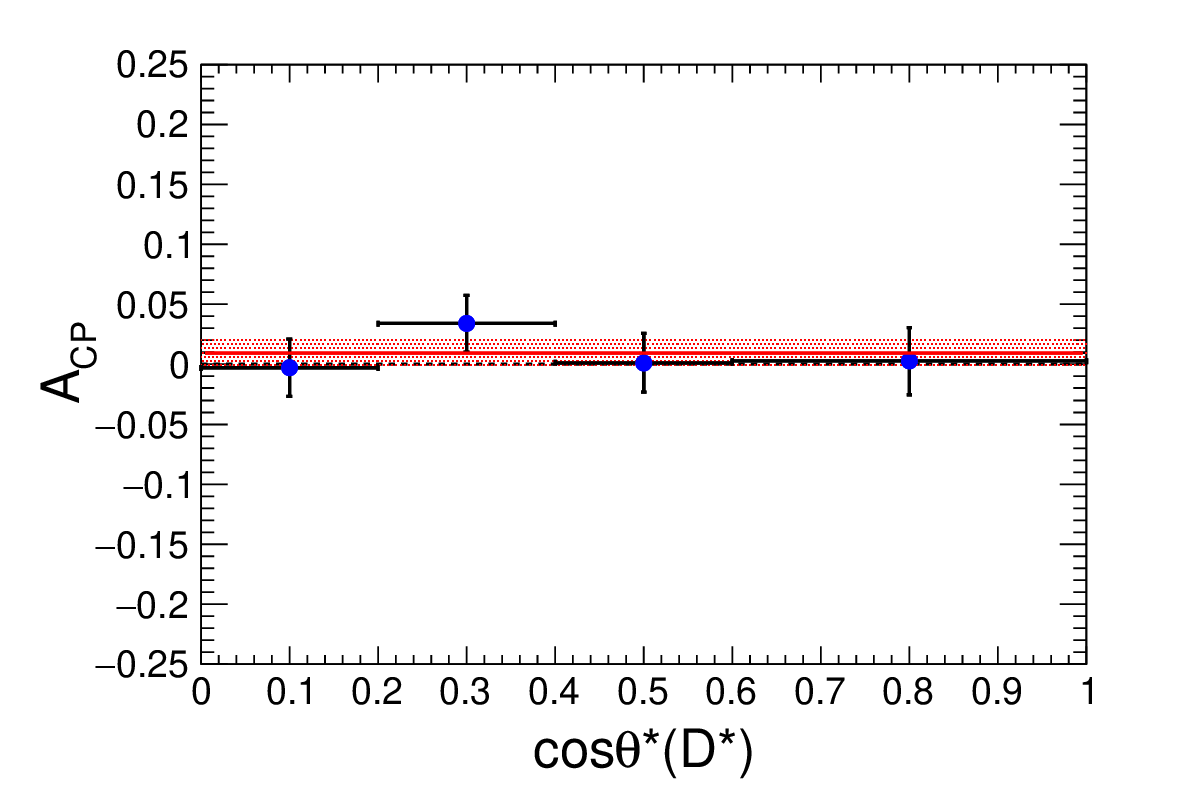}%
  \put(75,55){(a)}
  \end{overpic}%
  \begin{overpic}[width=0.333\textwidth]{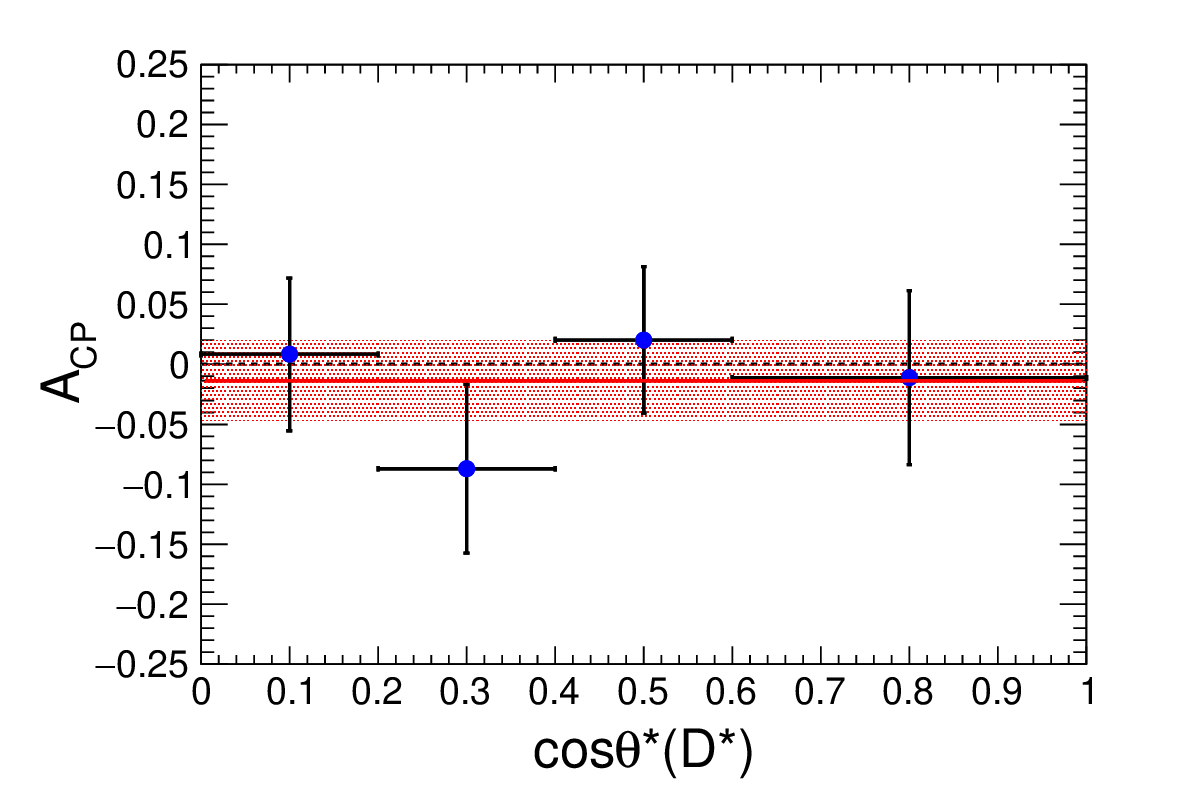}%
  \put(75,55){(b)}    
  \end{overpic}%
  \begin{overpic}[width=0.333\textwidth]{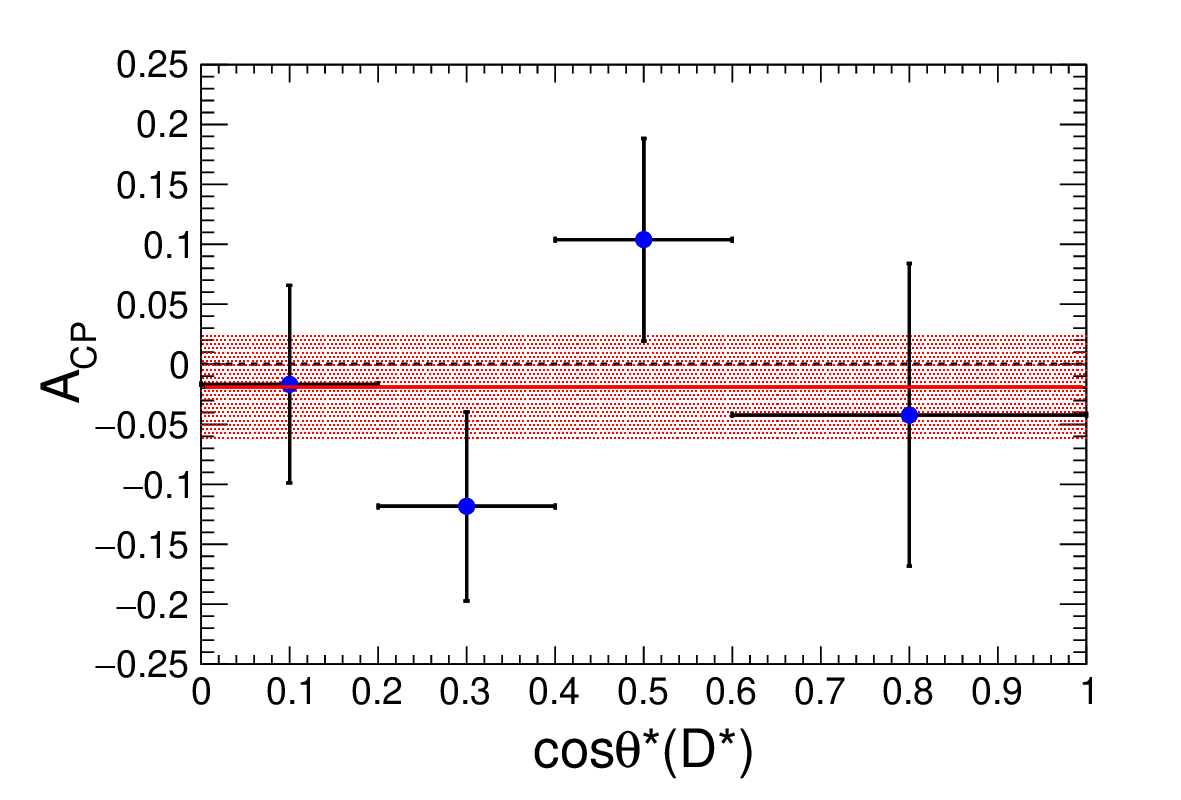}%
  \put(75,55){(c)}    
  \end{overpic} 
  \begin{overpic}[width=0.333\textwidth]{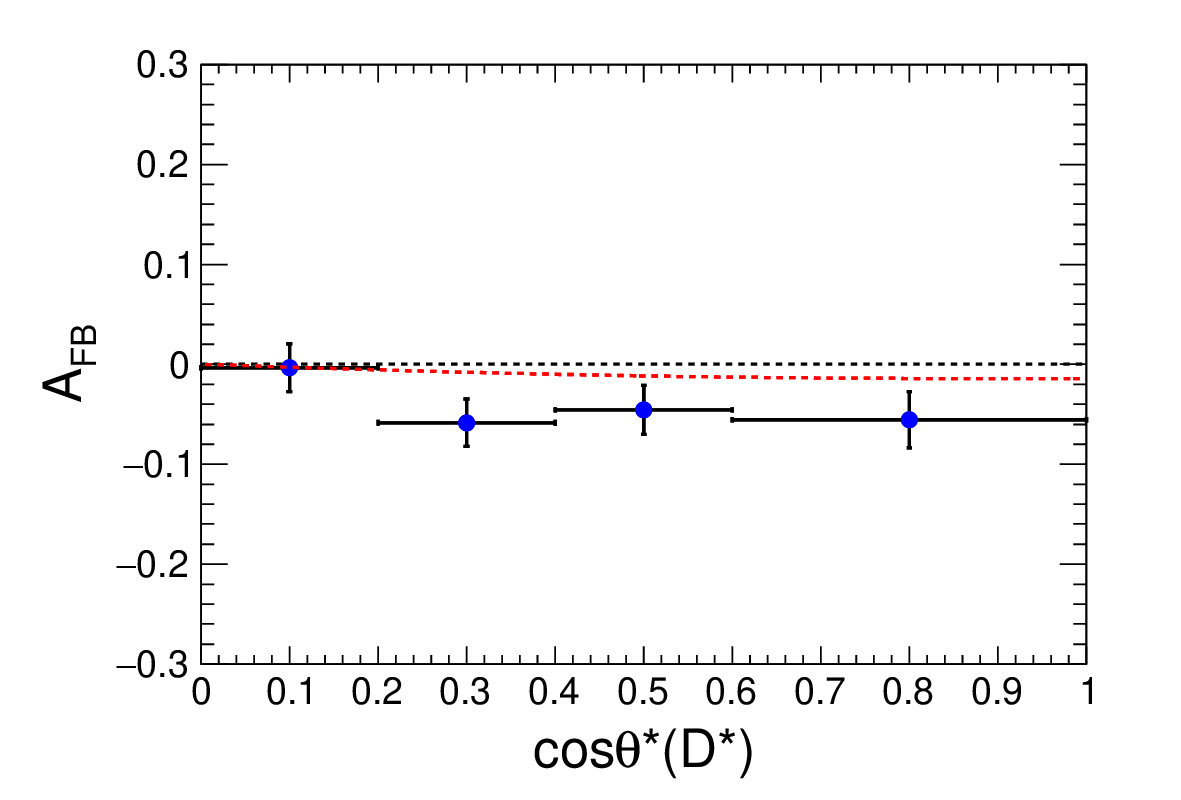}%
  \put(75,55){(d)}
  \end{overpic}%
  \begin{overpic}[width=0.333\textwidth]{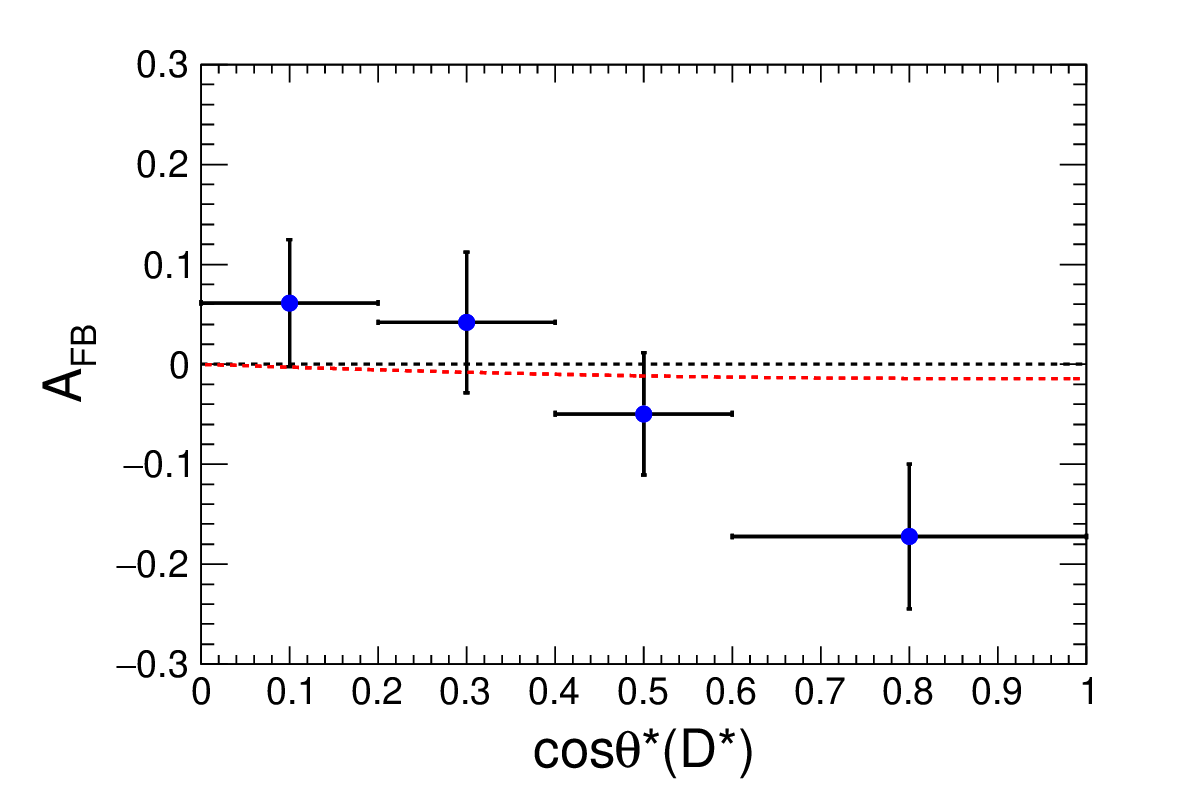}%
  \put(75,55){(e)}    
  \end{overpic}%
  \begin{overpic}[width=0.333\textwidth]{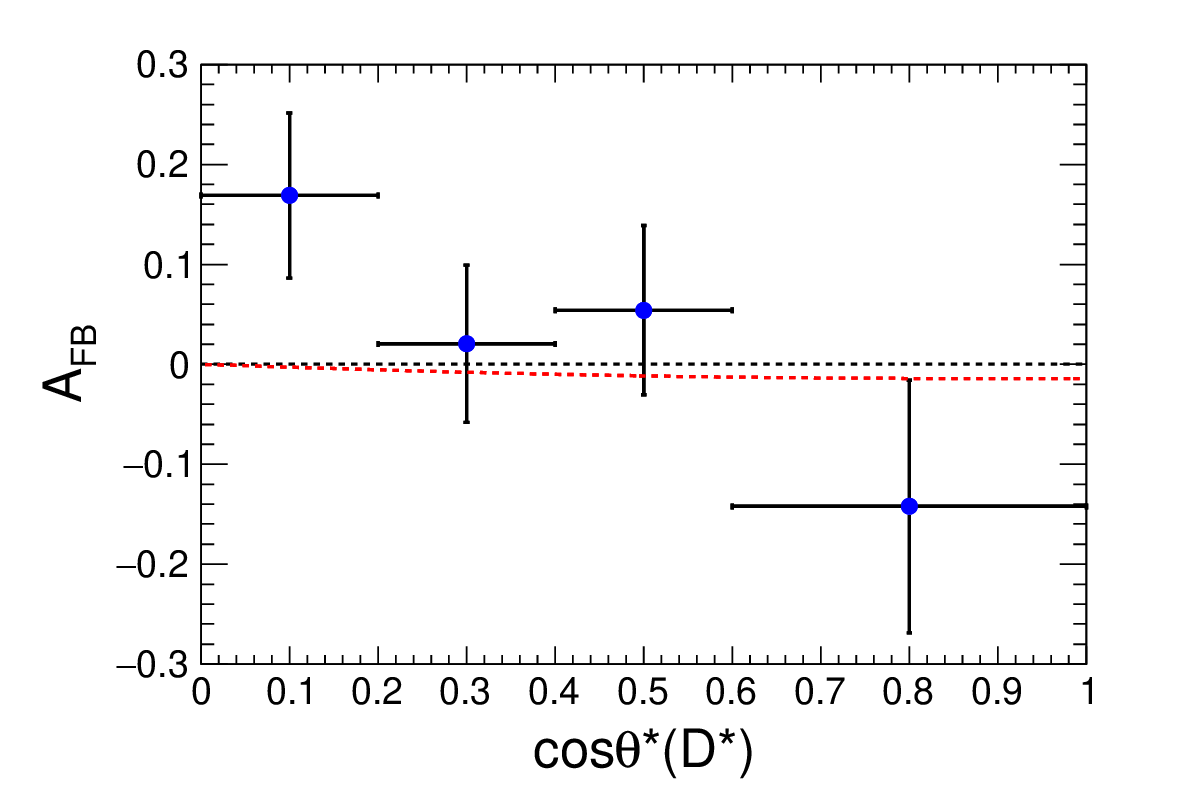}%
  \put(75,55){(f)}    
  \end{overpic}   
  \caption{\label{fig:Acp_cosTheta}$\CP$-violating asymmetry $\Acp$ (top) and forward-backward asymmetry $\Afb$ (bottom) values as a function of $\cos\theta^{*}(D^{*+})$ for (a, d) $\DzToPiPiEta$, (b, e) $\DzToKKEta$, and (c, f) $\DzToPhiEta$, respectively. The solid red lines with a band region are the averaged values with their uncertainties. The dashed red curves in the $\Afb$ plots show the leading-order prediction for $\Afb(\epem\to c\bar{c})$.}
  \end{center}
\end{figure}

We repeat the above procedure to determine $\Acp$ for $\DzToPhiEta$ decays. Here, to determine parameters 
$N_{\rm sig}$ and $A_{\rm corr}$, we perform a two-dimensional fit in $[M_{KK}, Q]$ 
for the $\Dz$ and $\Dzb$ samples simultaneously.
We allow $N_{\rm sig}$ and $A_{\rm corr}$
to float separately for the $\phi\eta$ and non-resonant $K^+K^-\eta$
components. 
The projections of the fit result are shown in 
figure~\ref{fig:unblindAcpDzToPhiEta}, and the results are
$N_{\rm sig}=728\pm 36$ and $A_{\rm corr}=(-0.17\pm 4.44)\%$.
We perform this fit separately to obtain the $A_{\rm corr}$ values for the eight bins of 
$\cos\theta^*$ and use eqs.~\eqref{eqn:Acp1}~and~\eqref{eqn:Acp2}~to 
extract $\Acp$ and $\Afb$. The resulting four values of $\Acp$ and $\Afb$ 
are plotted in figures~\ref{fig:Acp_cosTheta}(c,\,f). 
Fitting these $\Acp$ values to a constant gives
\begin{eqnarray}
A_{\CP}(\Dz\to\phi\eta) & = & [-1.9\pm 4.4\,({\rm stat})\pm 0.6\,({\rm syst})]\%\,,
\end{eqnarray}
where the second error listed is the systematic uncertainty, evaluated below (section~\ref{sec:systAcp}).
This result is consistent with zero, as expected~\cite{bib:PRD100d093002}. 

\subsection{Systematic uncertainties}
\label{sec:systAcp}
Fortunately, most systematic uncertainties in measuring 
$\Acp$ cancel. The remaining sources of systematic uncertainty 
are listed in table~\ref{tab:sysAcp} and are evaluated as follows.

\begin{table}[!htbp]
\begin{center}
\begin{tabular}{|c|c|c|c|} \hline  
Sources 			&  $\sigma_{\Acp}(\DzToPiPiEta)$ & $\sigma_{\Acp}(\DzToKKEta)$  	& $\sigma_{\Acp}(\DzToPhiEta)$ \\  \hline
Signal and bkg	    &  	$0.004$	&	$0.010$	&	$0.006$	\\ 
$\cos\theta^*$ binning    	
                    &  	$0.002$	&	$0.004$	&	$0.002$	\\  
$A_{\varepsilon}(\pi_s)$ map 
					&	$0.001$	&	$0.001$	&	$0.001$		\\ \hline 
Total syst. error	&  	$0.005$	&	$0.011$		& 	$0.006$		\\ \hline 
\end{tabular}
\caption{The absolute systematic uncertainties for $\Acp$ measurement in each SCS decay mode.}
\label{tab:sysAcp}
\end{center}  
\end{table}

\begin{itemize}
\item 
There is an uncertainty arising from fixed parameters in the fit used 
to describe signal and background shapes. We evaluate this uncertainty
using the sampling method described previously to evaluate  
uncertainties for the branching fraction measurement. The 
resulting uncertainties
for $\Acp$ are small: $<0.001$ for both $\DzToPiPiEta$ and 
$\DzToKKEta$, and $0.002$ for $\DzToPhiEta$. 
We also consider different possible $Q$ and $M_{KK}$ resolutions for the
$\Dz$ and $\Dzb$ samples by allowing the $\sigma^{}_0$ parameter
in eqs.~(\ref{eqn:qshape1}\,-\,\ref{eqn:qshape3}) and 
the $\sigma_{m0}$ and $\sigma^{}_{0}$ parameters in eqs.~(\ref{eqn:MphiQ2D_sig0},\,\ref{eqn:MphiQ2D_sig1})
to vary between the two samples. The change in $\Acp$ is 
$0.004$ for $\DzToPiPiEta$, $0.010$ for $\DzToKKEta$, and $0.006$ 
for $\DzToPhiEta$. Combining these two uncertainties in quadrature
gives the values listed in table~\ref{tab:sysAcp}.

\item 
We extract $\Acp$ via a binning procedure in $\cos\theta^*$ 
[see eqs.~\eqref{eqn:Acp1}-\eqref{eqn:Acp2}], and 
there is possible uncertainty arising from the choice of bins used.
We thus change the number of bins from eight to six, with
bin divisions ($-1.0$, $-0.55$, $-0.27$, $0.0$, $0.27$, $0.55$, $1.0$). 
The resulting change in $\Acp$ is taken as the systematic uncertainty
due to this source.
There is a small uncertainty arising from a difference in 
the detector acceptance near the boundaries $\cos\theta^*=\pm 1$;
we evaluate this by considering only events with $|\cos\theta^*|<0.90$.

\item 
We correct for a possible asymmetry in $\pi^\pm_s$
reconstruction efficiencies by weighting events according
to a mapping of efficiencies $\varepsilon(\pi_s)$. There
are 56 bins in this map, and the efficiencies for each
bin has some uncertainty. We thus vary 
these efficiencies individually
by their uncertainties to create 56 new
efficiency maps with $+1\sigma$ shifts and 56 maps with
$-1\sigma$ shifts. We subsequently weight the $\Dz$ and
$\Dzb$ samples by these efficiency maps and repeat
the fit for $N^{}_{\rm sig}$ and $A_{\CP}^{\rm cor}$.
The resulting deviations from the nominal fit result are 
summed in quadrature to give the systematic uncertainty
arising from this source. We obtain ${}^{+0.050}_{-0.087}\%$ for $\DzToKKEta$, ${}^{+0.065}_{-0.072}\%$ for $\DzToKKEta$, and ${}^{+0.043}_{-0.100}\%$ for $\DzToPhiEta$. 

\end{itemize}
The total systematic uncertainty is obtained by adding 
in quadrature all the above contributions. The results
are listed in table~\ref{tab:sysAcp}. 

\section{Conclusion}\label{sec:conclusion}
In summary, based on a data set
corresponding to an integrated luminosity of
980 $\rm fb^{-1}$ recorded by the Belle experiment,
we report measurements of the branching fractions 
of the SCS decays $\DzToPiPiEta$ and 
$\DzToKKEta$ relative to that for the CF decay $\DzToKPiEta$.
We also measure the relative branching fraction for the resonant decay $\DzToPhiEta$; this measurement uses an order of magnitude more data 
than used for our previous measurement~\cite{bib:PRL92d101803} and
supersedes it. Our results are:
\begin{eqnarray}
 \frac{\mathcal{B}(\DzToPiPiEta)}{\mathcal{B}(\DzToKPiEta)} & = & 
 [6.49 \pm 0.09\,({\rm stat})\pm 0.13\,({\rm syst})]\%\,, \\ 
 \frac{\mathcal{B}(\DzToKKEta)}{\mathcal{B}(\DzToKPiEta)} & = &
 [0.957\,^{+0.036}_{-0.033}\,({\rm stat})\pm 0.021\,({\rm syst})]\%\,, \\
 \frac{\mathcal{B}(D^0\to\phi\eta,\phi\to\Kp\Km)}
                {\mathcal{B}(\Dz\to\Km\pip\eta)}   & = &  
 [4.82\pm 0.23\,({\rm stat})\pm 0.16\,({\rm syst})]\times 10^{-3}\,.
 \end{eqnarray}
The color-suppressed decay $\DzToPhiEta$ is observed for the first time, with high statistical significance. Multiplying the above results by the world average value 
$\mathcal{B}(\DzToKPiEta)=(1.88\pm 0.05)\%$~\cite{bib:PDG2020}
gives the following absolute branching fractions:
\begin{eqnarray}
\mathcal{B}(\DzToPiPiEta) & = & 
[1.22\pm 0.02\,({\rm stat})\pm 0.02\,({\rm syst}) \pm 0.03\, (\mathcal{B}_{\rm ref}) ]\times 10^{-3}\,, \\
\mathcal{B}(\DzToKKEta) & = & 
[1.80\,^{+0.07}_{-0.06}\,({\rm stat})\pm 0.04\,({\rm syst}) \pm 0.05\, (\mathcal{B}_{\rm ref}) ]\times 10^{-4}\,,  \\
\mathcal{B}(\Dz\to\phi\eta) & = & 
[1.84\pm 0.09\,({\rm stat})\pm 0.06\,({\rm syst}) \pm 0.05\, (\mathcal{B}_{\rm ref}) ]\times 10^{-4}\,, 
\end{eqnarray}
where the third uncertainty is due to the branching fraction for
the reference mode $\Dz\to\Km\pip\eta$.
These results are the most precise to date. 

The time-integrated $\CP$ asymmetries are measured to be
\begin{eqnarray}
A_{\CP}(\DzToPiPiEta) & = & 
[0.9\pm 1.2\,({\rm stat})\pm 0.4\,({\rm syst})]\%\,,  \\
A_{\CP}(\DzToKKEta) & = & 
[-1.4\pm 3.3\,({\rm stat})\pm 1.0\,({\rm syst})]\%\,, \\
A_{\CP}(\Dz\to\phi\eta) & = & 
[-1.9\pm 4.4\,({\rm stat})\pm 0.6\,({\rm syst})]\%\,. 
\end{eqnarray}
The first result represents a significant improvement in precision 
over the previous result~\cite{bib:PRD101d052009}. The latter
two are the first such measurements. No evidence for $\CP$
violation is found.

\acknowledgments
We thank the KEKB group for the excellent operation of the
accelerator; the KEK cryogenics group for the efficient
operation of the solenoid; and the KEK computer group, and the Pacific Northwest National
Laboratory (PNNL) Environmental Molecular Sciences Laboratory (EMSL)
computing group for strong computing support; and the National
Institute of Informatics, and Science Information NETwork 5 (SINET5) for
valuable network support.  We acknowledge support from
the Ministry of Education, Culture, Sports, Science, and
Technology (MEXT) of Japan, the Japan Society for the 
Promotion of Science (JSPS), and the Tau-Lepton Physics 
Research Center of Nagoya University; 
the Australian Research Council including grants
DP180102629, 
DP170102389, 
DP170102204, 
DP150103061, 
FT130100303; 
Austrian Federal Ministry of Education, Science and Research (FWF) and
FWF Austrian Science Fund No.~P~31361-N36;
the National Natural Science Foundation of China under Contracts
No.~11435013,  
No.~11475187,  
No.~11521505,  
No.~11575017,  
No.~11675166,  
No.~11705209;  
Key Research Program of Frontier Sciences, Chinese Academy of Sciences (CAS), Grant No.~QYZDJ-SSW-SLH011; 
the  CAS Center for Excellence in Particle Physics (CCEPP); 
the Shanghai Pujiang Program under Grant No.~18PJ1401000;  
the Shanghai Science and Technology Committee (STCSM) under Grant No.~19ZR1403000; 
the Ministry of Education, Youth and Sports of the Czech
Republic under Contract No.~LTT17020;
Horizon 2020 ERC Advanced Grant No.~884719 and ERC Starting Grant No.~947006 ``InterLeptons'' (European Union);
the Carl Zeiss Foundation, the Deutsche Forschungsgemeinschaft, the
Excellence Cluster Universe, and the VolkswagenStiftung;
the Department of Atomic Energy (Project Identification No. RTI 4002) and the Department of Science and Technology of India; 
the Istituto Nazionale di Fisica Nucleare of Italy; 
National Research Foundation (NRF) of Korea Grant
Nos.~2016R1\-D1A1B\-01010135, 2016R1\-D1A1B\-02012900, 2018R1\-A2B\-3003643,
2018R1\-A6A1A\-06024970, 2018R1\-D1A1B\-07047294, 2019K1\-A3A7A\-09033840,
2019R1\-I1A3A\-01058933;
Radiation Science Research Institute, Foreign Large-size Research Facility Application Supporting project, the Global Science Experimental Data Hub Center of the Korea Institute of Science and Technology Information and KREONET/GLORIAD;
the Polish Ministry of Science and Higher Education and 
the National Science Center;
the Ministry of Science and Higher Education of the Russian Federation, Agreement 14.W03.31.0026, 
and the HSE University Basic Research Program, Moscow; 
University of Tabuk research grants
S-1440-0321, S-0256-1438, and S-0280-1439 (Saudi Arabia);
the Slovenian Research Agency Grant Nos. J1-9124 and P1-0135;
Ikerbasque, Basque Foundation for Science, Spain;
the Swiss National Science Foundation; 
the Ministry of Education and the Ministry of Science and Technology of Taiwan;
and the United States Department of Energy and the National Science Foundation.

\appendix
\section{Bifurcated Student's t-function\label{app:student}}
The bifurcated Student's t-function is defined as: 
\begin{eqnarray}\label{eqn:bifStu}
  S_{\rm bif}(x; \mu, \sigma, \delta, n_l, n_h)&=& \frac{2P_H P_L}{(P_H+P_L)\sqrt{\pi}}
  \left\{
    \begin{array}{ll}
      \left[1+\frac{1}{n_h}\left(\frac{x-\mu}{\sigma(1+\delta)}\right)^2\right]^{-\frac{n_h+1}{2}}, & \hbox{for $x\geq\mu$;} \\
      \left[1+\frac{1}{n_l}\left(\frac{x-\mu}{\sigma(1-\delta)}\right)^2\right]^{-\frac{n_l+1}{2}}, & \hbox{for others.}
    \end{array}
  \right. 
\end{eqnarray}
Here the factors $P_H$ and $P_L$ are calculated as:
$P_H=\frac{\Gamma(\frac{n_h+1}{2})}{\sigma\cdot(1+\delta)\Gamma(\frac{n_h}{2})}\frac{1}{\sqrt{n_h}}$ and
$P_L=\frac{\Gamma(\frac{n_l+1}{2})}{\sigma\cdot(1-\delta)\Gamma(\frac{n_l}{2})}\frac{1}{\sqrt{n_l}}$,
where $\Gamma$ is the Gamma function.

%
\bibliographystyle{JHEP.bst}
\bibliography{references.bib}


\end{document}